\documentclass{aa}
\usepackage{graphicx}
\usepackage{xcolor}
\usepackage{caption}
\usepackage{subcaption}
\usepackage{txfonts}
\usepackage{ulem}
\def\Lsun{L$_{\odot}$}
\def\Msun{M$_{\odot}$}
\def\msun{M$_{\odot}$}

\def\Rsun{R$_{\odot}$}
\defcitealias{2020A&A...640A..15R}{Paper~I}
\begin{document}

   \title{The Carina High-contrast Imaging Project for massive Stars (CHIPS)}
   \subtitle{II. O stars in Trumpler~14}

   \author{A. Rainot\inst{1,2}, M. Reggiani\inst{1}, H. Sana\inst{1}, J. Bodensteiner\inst{1,3}, O.~Absil\inst{4}\fnmsep\thanks{F.R.S.-FNRS Senior Research Associate}}
   \institute{Institute of Astronomy, KU Leuven, Celestijnenlaan 200D, 3001 Leuven, Belgium\\
              \email{alan.rainot@kuleuven.be}
         \and
              ScanWorld SA, Rue des Chasseurs Ardennais 6, Liège Science Park, 4031 Angleur, Belgium
         \and
              European Southern Observatory, Karl-Schwarzschild-Str. 2, 85748 Garching, Germany
         \and
              Space sciences, Technologies and Astrophysics Research (STAR) Institute, Universit\'e de Li\`ege, 19 All\'ee du Six Ao\^ut, 4000 Li\`ege, Belgium
            }

   \titlerunning{CHIPS II. O stars in Trumpler~14}
   \authorrunning{A.~Rainot et al.}
   
\date{Accepted 22.11.2021}

\abstract{
      Most massive stars belong to multiple systems, yet the formation process leading to such high multiplicity remain insufficiently understood. 
     To help constrain the different formation scenarios that exist, insights on the low-mass end of the companion mass function of such stars is crucial. However, this is a challenging endeavour as (sub-)solar mass companions at angular separations ($\rho$) below 1\arcsec\ (corresponding to 1000–3000~au in nearby young open clusters and OB associations) are difficult to detect due to the large brightness contrast with the central star. 
}{
    With the Carina High-contrast Imaging Project of massive Stars (CHIPS), we aim to obtain statistically significant constraints on the presence and properties of low-mass companions around massive stars at a previously unreachable observing window ($\Delta \mathrm{mag} \gtrsim 10$ at $\rho \lesssim 1\arcsec$). In this second paper in the series, we focus on the Trumpler~14 cluster, which harbours some of the youngest and most massive O-type stars in the Milky Way. 
}{
    We obtained VLT-SPHERE observations of seven O-type objects in Trumpler~14 using the IRDIFS\_EXT mode. These provide us with a $12\arcsec \times 12\arcsec$ field-of-view (approximately, (($3 \times 3)  \cdot 10^4$~au) centred on each O star and allow us to search for companions at separations larger than 0\farcs15 (approx.\ 360~au) and down to magnitude contrast $>10~\mathrm{mag}$ in the near-infrared. We used  angular and spectral differential imaging along with PSF fitting to detect sources and measure their flux relative to that of the central object. We then used grids of ATLAS9 and PHOENIX LTE atmosphere models combined with (pre-) main-sequence evolutionary tracks to estimate the mass of the detected candidate companions. 
}{We detected 211 sources with near-infrared magnitude contrast in the range of 2 to 12. Given the large surface number density of stars in Trumpler 14, one cannot reliably distinguish between cluster members and genuine companions for most of the detected sources. The closest companion, at only 0\farcs26, is characterised as a 1.4~\msun{} stars with an age of 0.6~Myr, in excellent agreement with previous age estimates for Tr~14. The mass function peaks at about 0.4~\Msun{} and presents a dearth of stars in the 0.5 to 0.8~\Msun{} mass range compared to previous estimates of the initial mass function in Tr 14. While statistically significant, part of these differences may result from contamination of the K-band fluxes by circumstellar material.
}{
SPHERE is clearly suitable to probe the low-mass end of the mass function in the vicinity of massive stars. Follow-up SPHERE observations to obtain the full Y to K spectral energy distribution would allow for better constraints on the masses of the detected sources, and to confirm (or invalidate) the curious mass function that we derived for low-mass stars in the vicinity of the O-type objects in Trumpler 14.
}

\maketitle

\section{Introduction}\label{s:intro}

   Despite the importance of massive stars throughout the history of the Universe, massive star formation remains an important open question in modern astrophysics.    Several formation scenarios have been proposed to explain the formation of massive stars, including  stellar collisions \citep{1998MNRAS.298...93B}, competitive accretion \citep{2001MNRAS.323..785B,2006MNRAS.370..488B} and monolithic collapse \citep{2003ApJ...585..850M,2009Sci...323..754K}. Most theories agree on the presence of a dense and massive accretion disk to overcome the radiation barrier \citep{2010ApJ...722.1556K}. In massive stars, such disk is expected to fragment under gravitational instabilities \citep{2010ApJ...708.1585K}, possibly leading to clumps massive enough to form low-mass stars. While theoretical predictions for the properties of these clumps remain scarce, observational constraints on the low-mass end of the companion mass function of massive stars would certainly provide valuable clues on their formation process.
   Indeed, the influence of environmental factors, as for instance the stellar density or the feedback from massive stars, on the shape of the Initial Mass Function (IMF) at its low-mass end  remains an open question. While the low-mass and sub-stellar IMF has been extensively studied in nearby star forming regions \citep[e.g.][]{2016ApJ...827...52L,2018AJ....156..271L, 2019MNRAS.486.1718S}, only a limited number of studies have analysed dense clusters where massive stellar feedback dominates \citep{2016A&A...588L...7K}. A recent study by \cite{2021arXiv210108804D} has analysed the IMF of eight young (< 5~Myrs old) clusters with diverse stellar densities and frequency of massive stars and concluded that there is no significant evidence for environmental effects on the shape of the IMF in these regions.
        
    The Carina High-contrast Imaging Project of massive Stars (CHIPS) aims to investigate these questions by performing high-contrast observations of the immediate surroundings of O- and Wolf-Rayet (WR) type stars in the Carina region 
    using the SPHERE instrument \citep{2019A&A...631A.155B} at the ESO Very Large Telescope (VLT). SPHERE allows us to probe the close environment of massive stars in the angular separation range of $0\farcs15 < \rho \lesssim 6\farcs5$  with a contrast of up to twelve magnitudes between sources in the field of view (FoV) and the central object. This corresponds to projected separations in the range of approximately\ 375 to 16\,000~au  (2 to $80\times 10^{-3}$~pc) in the Carina region. The inner part of this separation range has never been probed to such contrasts and these observations will help to verify whether the plethora of faint companions discovered at separations larger than a few 1000 au \citep{2014ApJS..215...15S,2015AJ....149...26A,2020AJ....160..115C} also exists at shorter separations or whether there is a critical separation scale at which the distribution breaks down.
    
\begin{figure}
     \centering
        \includegraphics[width=.95\columnwidth]{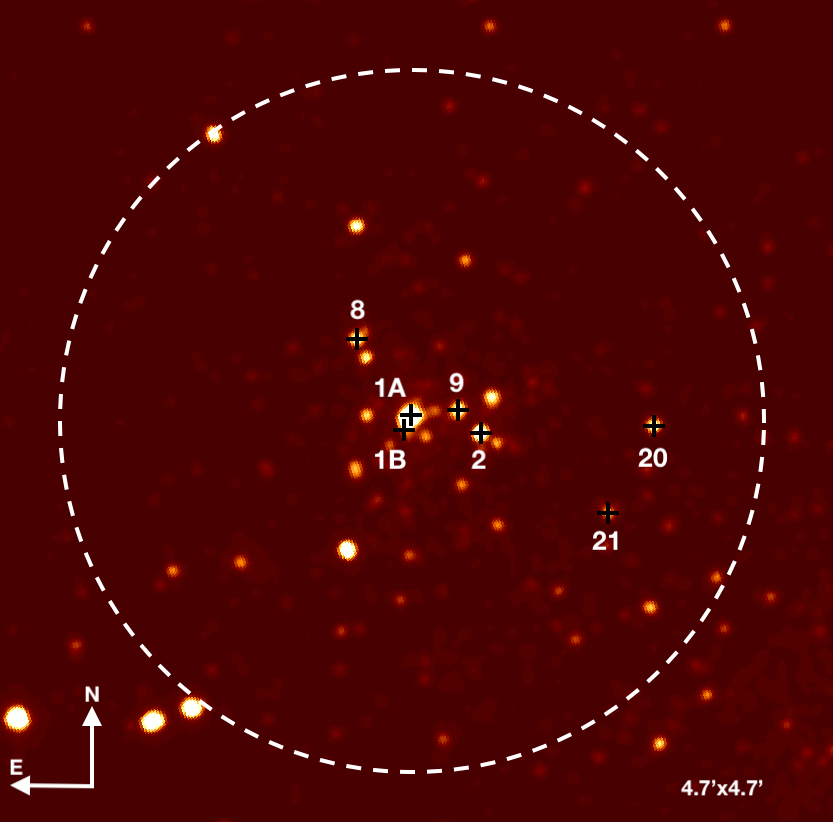}\hspace*{5mm}
      \caption{2MASS image of the Trumpler~14 cluster with its seven O-type stars and multiple systems indicated with crosses \citep{2006AJ....131.1163S}. The white circle shows the 2' area around the central star of the cluster, Tr14-1A, considered for this study.}
      \label{f:tr14}
   \end{figure}

    This present paper is the second entry in a short series. After establishing the proof-of-concept in the first paper on the QZ Car system \citep[][hereafter \citetalias{2020A&A...640A..15R}]{2020A&A...640A..15R} which found 19 sources including a close companion at $\sim$1700 au,  we now focus on the O-type stars in the Trumpler~14 cluster. Trumpler~14 (Tr~14) is a young massive cluster at the North end of the Carina region. At a distance of about 2.5~kpc from Earth \citep[]{2004A&A...418..525C, 2019ApJ...870...32K}, it is one of the densest and youngest massive star clusters in the solar neighbourhood (see Sect. \ref{s:phys_source}). It has an estimated age smaller than 1~Myr and contains seven O-type stars that are individually observable with SPHERE \citep[e.g.][]{2007A&A...476..199A,2010A&A...515A..26S, 2011MNRAS.418..949R}. This makes it an ideal environment to study the birth multiplicity of massive stars. Previous Adaptive Optics (AO) studies with the Multi-conjugate Adaptive optics Demonstrator \citep[MAD,][]{2010A&A...515A..26S} at the VLT have found 150 likely bound pairs with 30\%\ of them being close pairs at separations smaller than $0\farcs5$ ($\sim1300$ au). The authors have also showed that at the $2\sigma$ level, massive stars in Tr~14 have more companions than lower mass stars. These companions are more massive than the ones of low-mass stars  \citep{2010A&A...515A..26S}. Here, we observe all seven O-type objects that can be spatially resolved from each other with VLT/SPHERE, allowing us to map the cluster properties at various radial distances (Fig.~\ref{f:tr14}). 
    
    This paper is organised as follows. Sect.~\ref{s:data} presents the observations, data reduction and the image post-processing algorithms. Results are discussed in Sect.~\ref{s:results} and our conclusions are presented in Sect.~\ref{s:Ccl}.

\section{Observations and data reduction} \label{s:data}

  \subsection{Observations}\label{s:obs}
  
    Observations of the seven O-type systems in Tr~14  were obtained with VLT/SPHERE from April 2015 to March 2019  using the IRDIFS extended mode (IRDIFS\_EXT), which combines $YJH$-bands observations with the Integral Field Spectrograph \citep[IFS,][]{2008SPIE.7014E..3EC} and $K$-band observations with the Infra-Red Dual-beam Imaging and Spectroscopy \citep[IRDIS,][]{2008SPIE.7014E..3LD} sub-system. The IFS camera covers the inner $1\farcs73\times1\farcs73$ FoV with plate-scale of 7.46~milli-arcsec/pix (mas/pix) producing observing cubes of $290 \times 290$ pixels. Meanwhile, IRDIS has a larger $12\arcsec\ \times12\arcsec$ FoV with a 12.25~mas/pix plate-scale, yielding cubes of $1024 \times 1024$ pixels. 
    
     For each of our seven targets, the observing sequence was similar and consisted of a {\sc centre (C)} frame which is used to centre the coronagraph,  a {\sc flux (F)} image to characterise the Point Spread Function (PSF) of the central star, and finally an {\sc object (O)} series of integrations, where the central star's light is blocked by the coronagraph. This sequence was repeated three times in a row for each object. As the central stars are bright O-type stars, the neutral density filter ND\_1 was used for the {\sc flux} images to attenuate the incoming light, avoiding saturation of the detector. 
    
    Observations were obtained with the telescope in pupil-tracking mode, allowing us to record field rotations over a small range of parallactic angles.  The detector integration times (DIT) and number of DITs (NDIT) for the O and F observations are provided in Appendix~\ref{a:Obs} together with the atmospheric conditions and the parallactic angle variations (PAV). For each star, images in 41 wavelength channels were obtained covering the  $YJHK$ bands (39 channels in $0.95-1.35 \mu$m for IFS and 2 for IRDIS $2.11 - 2.251 \mu$m). The resulting science data is a four-dimensional (4D) cube consisting of  spatial dimensions, wavelength channel and parallactic rotation. The F PSF data is three dimensional (spatial dimensions and wavelength channels) and repeated three times in each observation sequence. The known properties of the O-type stars in Tr~14 are summarised in Table~\ref{table:1}.
     
    \begin{figure*}
     \centering
        \includegraphics[width=.99\columnwidth]{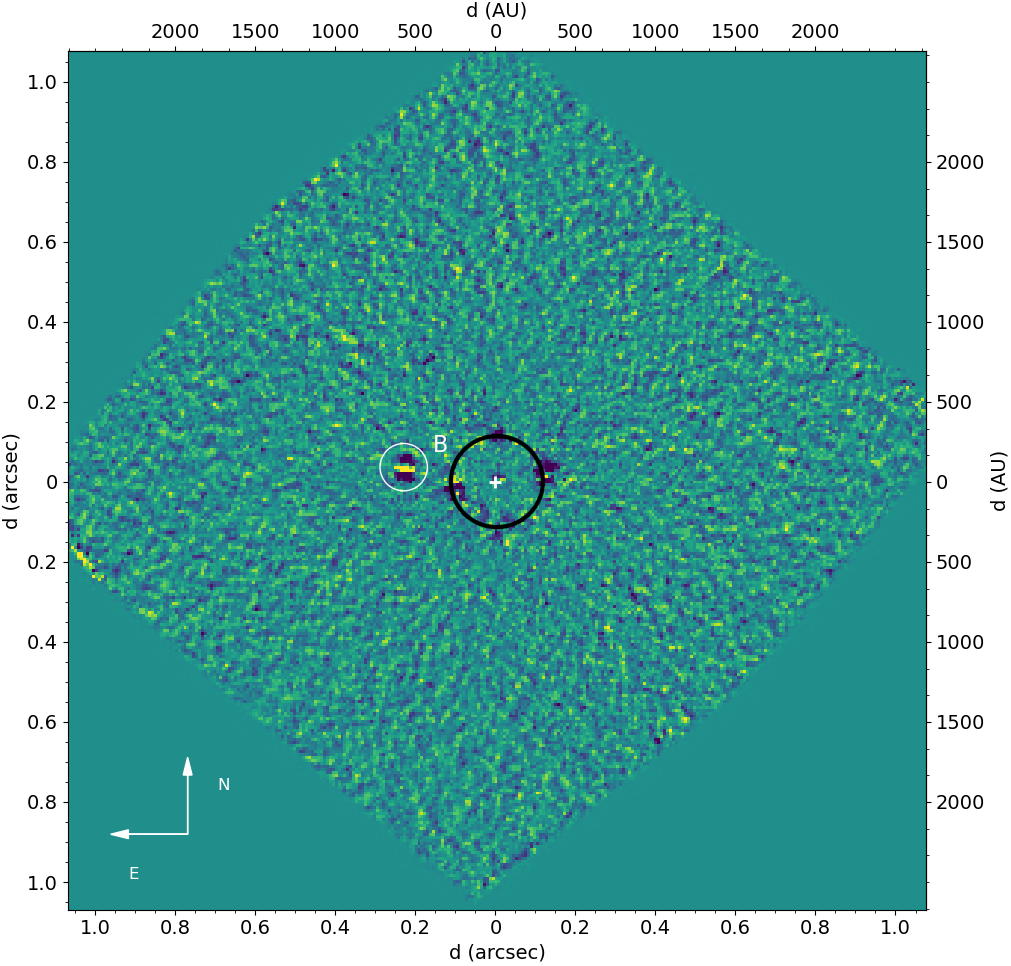}\hspace*{5mm}
        \includegraphics[width=.99\columnwidth]{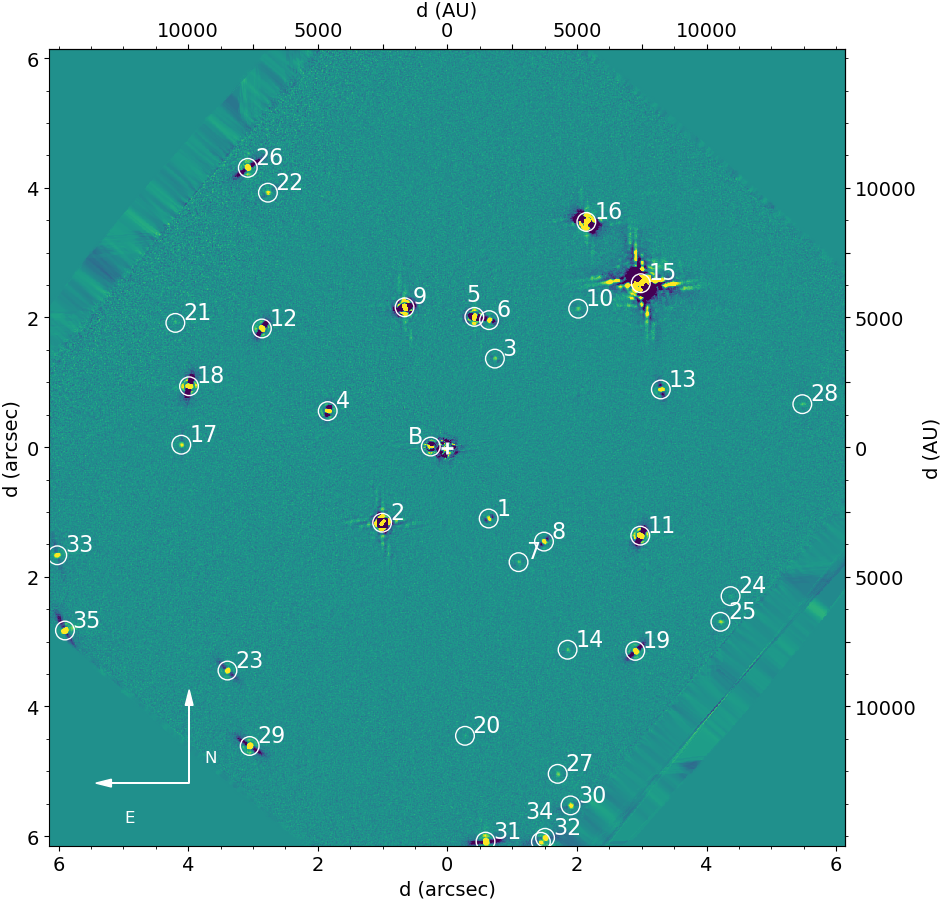}
      \caption{Post-processed collapsed images of the Tr14-8 FoV in the IFS (left) and IRDIS (right) observations ADI was applied using one principal component and one spectral channel (see Sect.~\ref{s:analysis}) for these images.       Dark circles on the centre of the images show the position of the $0\farcs173$-diameter coronagraph.  White circles indicate the sources detected at the $5\sigma$ level. The newly discovered 1.3~\msun\ (see Sect.~\ref{s:ifs}) companion Tr14-8B is clearly visible in the images. 
      } 
      \label{f:CD-583529}
    \end{figure*}
    
  \subsection{Science data reduction} \label{s:reduction}    
  
    The data reduction was performed by the SPHERE Data Center \citep[SPHERE-DC,][]{2017sf2a.conf..347D} at the Institut de Planetologie et d'Astrophysique de Grenoble (IPAG)\footnote{\url{http://ipag.osug.fr/?lang=en}}. Bad pixels, bias, dark and flatfield calibrations were applied for each exposure. The astrometry of science frames was corrected using the on-sky calibrations from \citet{2016SPIE.9908E..34M}, a True North correction value of $1.75\pm0.08^\circ$ and a plate scale of $7.46\pm0.02$ mas/pixel for IFS and $12.255\pm0.009$ mas/pixel for IRDIS. 4D science data cubes, tables including wavelengths and rotational angle information and the 3D PSF cubes are the products of the reduction.

    The three {\sc flux (F)} images taken of the central star without the coronagraph were median-averaged and subsequently used to compute the PSF and flux of the central star at each wavelength channel. A correction for the ND filter's transmission was applied during the data reduction. In this context, the IRDIS PSF images of three of our sources (Tr14-1B, Tr14-8 and Tr14-20) displayed shapes that were not representative of the true optical response of the instrument. For Tr14-8, the issue was related to the presence of a second source in the PSF image which interfered in the analysis. A mask was applied, where we replaced the impacted pixels with data  taken at the same radial distance but at a different position angle. For the  other two sources, the star was slightly off-centre and the PSF was deformed as a result, which would lead to artefacts in the analysis. For these sources, the reference PSF was
    replaced by that of Tr14-1A (HD~93129A) taken on the night of February 10, 2016, following the same procedure as described in \citetalias{2020A&A...640A..15R}. As noted in \citetalias{2020A&A...640A..15R}, Tr14-1A is a long period binary but it was not resolved at the time of the observations. The Tr14-1A PSF flux was scaled so that the integrated flux remained identical to that of each source's PSF. This allowed us to preserve the sources original flux information but with a PSF shape more in line with the expected behaviour of the instrument. 
        
    \begin{table*}
    \caption{Atmospheric and wind parameters adopted for the FASTWIND modelling of the SED of the central objects. Spectral types are taken from \citet{2013msao.confE.198M} with calibration tables from \citet{2005A&A...436.1049M}, apart from Tr14-1Aa and Tr14-1Ab for which parameters are taken from \citet{2019A&A...621A..63G}. Wind parameters were computed following \citet{2001A&A...369..574V}.}
    \label{table:1}
    \centering
    \begin{tabular}{l l c c c c c c c}
    \hline\hline
    \vspace*{-3mm}\\
    Star & Spectral  & $T_{\mathrm{eff}}$ & $\log g$ & $R_{\ast}$ & $M_{\ast}$ & $\log L_{\ast}$ & $\log\dot{M}$ & $v_{\infty}$ \\
     &Type & (K) & [cgs] & (R$_{\odot}$) & (M$_{\odot}$) & [L$_{\odot}$] &[M$_{\odot}$\,yr$^{-1}$]      & (km\,s$^{-1}$) \\
    \hline
       Tr14-1Aa & O2~I   & 52\,000 & 4.10 & 14.7 & 100   & 6.15 & $-4.70$ & 3200 \\
       Tr14-1Ab & O3~III & 45\,000 & 4.00 & 10.0 &  37   & 5.58 & $-5.80$ & 3200\\
       Tr14-1B  & O3.5~V & 44\,852 & 3.92 & 13.8 &  58 & 5.84 & $-5.35$ & 3355\\
       Tr14-2   & O3.5~V & 44\,852 & 3.92 & 13.8 &  58  & 5.84 & $-5.35$ & 3355\\
       Tr14-8   & O7~V   & 36\,872 & 3.92 &  9.2 &  25  & 5.14 & $-6.48$ & 2722\\
       Tr14-9   & O8.5~V & 33\,879 & 3.92 &  7.9 &  19  & 4.86 & $-7.00$ & 2525\\
       Tr14-20  & O6~V   & 38\,867 & 3.92 & 10.1 &  31  & 5.32 & $-6.16$ & 2866\\
       Tr14-21  & O9~V   & 32\,882 & 3.92 &  7.5 &  17  & 4.77 & $-7.18$ & 2466\\
    \hline
    \label{table:fastwind}
    \end{tabular}

    \end{table*}
    
  \subsection{Data analysis} \label{s:analysis}
 
    We followed the procedure detailed in \citetalias{2020A&A...640A..15R}. To summarise, we used the post-processing algorithm package Vortex Image Processing\footnote{\url{https://github.com/vortex-exoplanet/VIP}} \citep[VIP,][]{2017AJ....154....7G} to detect companions and extract their flux at each wavelength channel.  
    Different high-contrast imaging techniques were applied to reduce the data, such as angular differential imaging (ADI) and spectral differential imaging (SDI) using principal component analysis \citep[PCA,][]{2012MNRAS.427..948A,2012ApJ...755L..28S}. ADI makes use of the angular information from the parallactic angle variations while SDI utilises the wavelength data from each spectral channel to further suppress the speckles in the images. PCA is a method employed during the stacking of the different observing frames, enabling an increased Signal-to-Noise ratio (S/N), while suppressing speckles and reducing noise. For all of our targets, we used one principal component as it provided the necessary suppression of speckles and the self-subtraction of the sources was not prominent. The data analysis was streamlined in a pipeline, the Sphere High-contrast Imaging Pipeline for massive Stars (SHIPS), which is publicly released for reproducibility and possible further use by other teams\footnote{\url{https://github.com/arainot/SHIPS}}. 
        
    The detection process consisted in a first visual inspection of the IFS and IRDIS post-processed images for sources, followed by the computation of a S/N map in order to evaluate whether the visually identified sources are reliable detections. As in \citetalias{2020A&A...640A..15R}, we set our detection limit at $S/N = 5$. Angular separations ($\rho$), position angles (PA), flux (or magnitude) contrasts for each wavelength channel were then fit for every source with respect to their central star. The contrast spectra (i.e., the relative flux as a function of the wavelength channel) were obtained  by using one or both of the following methods: a Simplex Nelder-Mead optimisation contained in VIP (hereafter referred to as the Simplex method) which applies a negative fake companion technique to retrieve the position and flux parameters, and a PSF fitting technique that extracts the same parameters using PSF modelling \citep{2020A&A...634A..51B} described in \citetalias{2020A&A...640A..15R}. The latter is useful as, beyond approximately 2", the central star's influence (i.e., its contribution to the photon noise) is limited and the background noise dominates. Simplex is therefore not required to efficiently extract precise source parameters and a more widely used PSF-fitting technique can be used to provide accurate astrometry and photometry of sufficiently bright companions \citep[][]{2020A&A...634A..51B,2020A&A...640A..15R}. Simplex remains however required for the faintest sources while PSF-fitting is further useful close to the detector edges. Errors were computed using a Monte-Carlo approach with Simplex which consisted in injecting a number of artificial sources (20 in our case) at the same radial distance and flux of a given source. Parallactic angles are varied for each injected source. 
    
    Using Simplex, flux and positions are measured and compared to the input values. Finally, the errors associated to these parameters for each source are the standard deviations computed at $1\sigma$. An example of the resulting post-processed image for the star Tr~14-8 is provided in Fig.~\ref{f:CD-583529} while all other images are provided in Appendix~\ref{a:SIRDIS_obs}. Results are summarised in Sect~\ref{s:sumobs} and the catalogue of all detections is provided in an online table. Fig.~\ref{f:CD-583529} shows the clear detection of a close companion in the IFS image, named Tr~14-8B.

    Estimation of the absolute fluxes of the sources (S) detected   in the coronagraphic images would require a $YJHK$-flux calibrated spectrum of the central objects. Unfortunately, no such spectra are available in the needed spectral range and we therefore modelled the spectral energy distribution (SED) of the central objects using the non-local thermodynamic equilibrium (non-LTE) atmosphere code FASTWIND \citep[][]{2005A&A...435..669P,2011A&A...536A..58R, 2018A&A...619A..59S}. If a central object has multiple known but unresolved components (as e.g., QZ~Car in \citetalias{2020A&A...640A..15R}), each component was modelled separately and the individual SEDs were added together to mimic the SED of the unresolved central system. Atmospheric parameters for these computation were adopted from the spectral-type calibration of \citet{2005A&A...436.1049M} using the spectral types from \citet{2013msao.confE.198M}, except for Tr14-1A for which we used the more detailed study of \citet{2019A&A...621A..63G}. An overview of the adopted parameters are given in Table~\ref{table:fastwind}, where the values for the mass-loss rate ($\dot{M}$) and terminal wind velocity ($v_\infty$) were computed following \citet{2001A&A...369..574V}.  With the model spectrum of each star computed, we obtained the absolute flux of sources by multiplying their contrast spectrum with the modelled SED of the central source.
    
 To check for the validity of our flux calibration process, we compared the IRDIS $K$-band magnitudes that we obtained to the $K_\mathrm{S}$ values reported by \citep{2010A&A...515A..26S} for overlapping sources in both studies. We adopt a cross-correlation radius of 0\farcs1. For all targets but Tr14-1A and 1B (HD~93129A and B) the mean difference was below or of the same order of magnitudes as the root mean squared (rms) dispersion between the SPHERE and MAD $K$-band photometry (typical rms values were of about 0.3~mag). Systematic shifts of 1.3 and 1.8~mag were however observed for Tr14-1A and 1B, respectively. These probably result from either the observational difficulties described above or from a less accurate representation of the adopted physical parameter vs. spectral-type calibration for such early O stars. To account for these systematic difference, we applied the above correction offsets to the K-band SPHERE photometry that we derived for these two objects.

    VIP also enabled us to compute the detection limits, or limiting contrast curves, of our observations. These contrast curves represent the maximum magnitude difference reachable at any given angular separation $\rho$ in the image. To avoid any conflict with the presence of companions on the images, we applied masks on all sources, which dramatically improve the shape of the resulting limiting contrast curves, although small artefacts remained and are visible in Fig.~\ref{f:contrast}. The median 5-$\sigma$ detection limit reaches  contrasts of $\Delta K1\approx 10$ and $\Delta K2 \approx 9$ at $0\farcs2$.

  \subsection{Companion detections} \label{s:sumobs}
   
    Throughout all seven observed FoVs, we detected 211 individual sources that we characterised to obtain their positions, magnitude contrast  and probability of spurious association. In this section, we provide more details for each observation. We cross-identified our detections with  previously known companions, in which case we preserved the original names of these companions. New sources detected with IRDIS that are not detected within IFS' FoV are referred to as `S' sources in the following. The list of detected sources as well as their properties are given in Table~2 which is available in electronic form at the CDS\footnote{https://cds.u-strasbg.fr}.
    \addtocounter{table}{+1}
    \subsubsection{IFS detection} \label{s:ifs}
    
\begin{figure}
      \centering
      \includegraphics[width=\hsize]{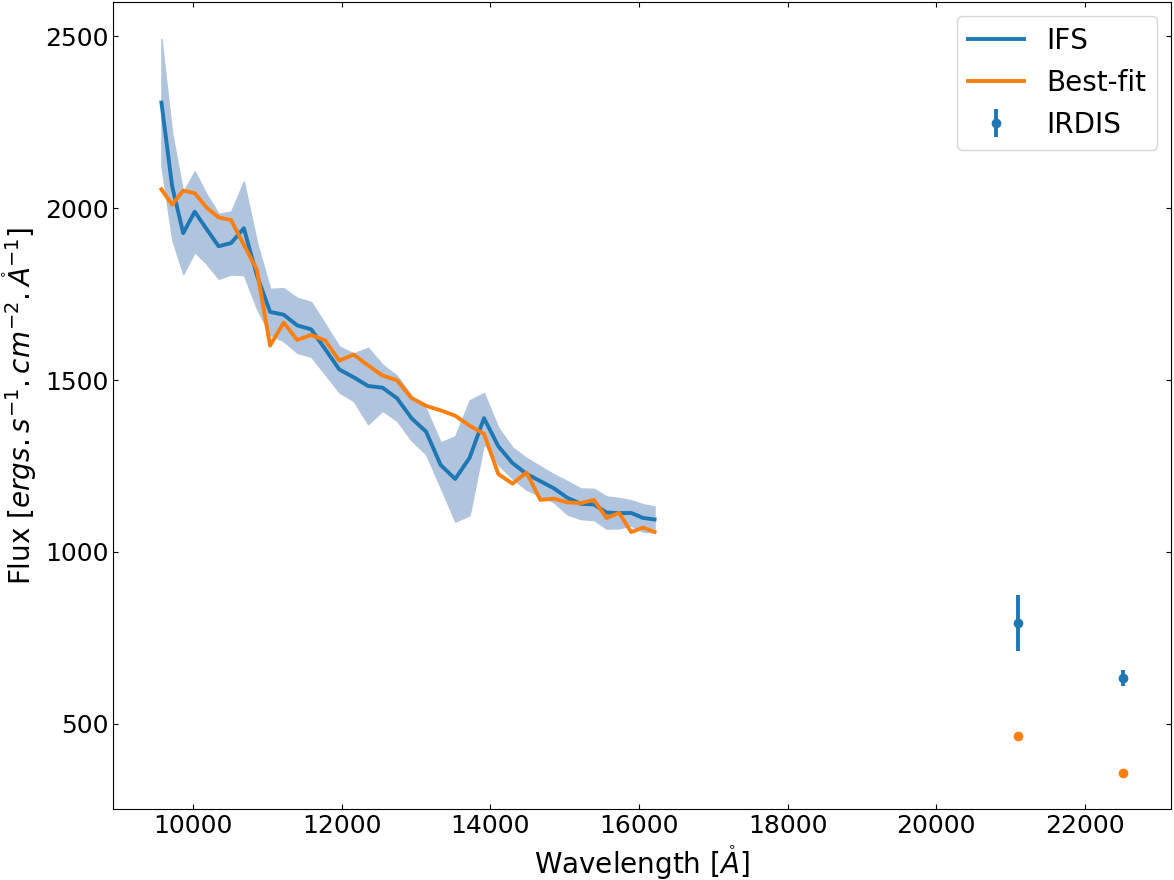}
         \caption{IFS+IRDIS flux-calibrated spectrum (blue) of Tr14-8B at a reference distance of 100~R$_\odot$. The orange plain line and points give the best-fit ATLAS9 model with $T_\mathrm{eff}=4455$~K, $\log g=3.4$ and $R=3.8$~R$_\odot$ for IFS and IRDIS. The shaded area represents the 1-$\sigma$ uncertainties on the observed spectrum.}
      \label{f:sed_CD-583529}
      \end{figure}

      Only one source was found in the IFS FoV. This previously undetected companion, labelled Tr14-8B (Fig.~\ref{f:CD-583529}), was detected at a radial distance of $0\farcs402 \pm 0\farcs 001$ ($1006.0 \pm 4.0$ au), a position angle of $81.0\pm0.1^{\circ}$ and has a median  magnitude contrast of $\Delta m_\mathrm{YJH} = 5.40 \pm 0.03$ in IFS and $\Delta m_\mathrm{K1,2} = 3.92 \pm 0.04$ in IRDIS. The flux-calibrated SED   is displayed in Fig.~\ref{f:sed_CD-583529}. 
      
      Stellar parameters are  estimated by $\chi^2$-minimisation using a grid of ATLAS9 LTE atmosphere models \citet{2003IAUS..210P.A20C} attached to pre-main sequence (PMS) evolutionary tracks of \citet{2000A&A...358..593S}. Taking the errors from the Simplex MC method, the best-fit $\chi^2$ was found to provide an acceptable fit when restricting the fit to the sole IFS spectrum, but not while including the two IRDIS channels. Flux values within those channels are significantly larger than expected from the best-fit IFS models, as seen in Fig.~\ref{f:sed_CD-583529}. Such  infrared excess could be due to the presence of a dusty disk around the companion, which is not unexpected given the young age of Trumpler~14. The best-fit model was obtained for a PMS star with $T_\mathrm{eff}=4455$~K, $R=3.8$~\Rsun{}, $\log g = 3.4 $, $L = 5.85$~\Lsun, $M = 1.4 M_{\odot}$ and an age of $6 \times 10^{5}$ years, which would correspond to a K5 star according to the calibration of \citet{2000A&A...358..593S}. The best fit is shown alongside the calibrated spectrum in Fig.~\ref{f:sed_CD-583529}. Our age estimate concurs with the age determined by \citet{2010A&A...515A..26S} for Trumpler~14. 
      
      Another source was visible in the IFS FoV of Tr14-9, however the source lays at the edge of the FoV and could only be seen for seven field rotations \textit{(i.e. seven IFS frames out of 24 in the cube}. As described in Sect.~\ref{s:irdis}, this companion is very close to another bright source, which resulted in speckles crossing the source flux so that we could not retrieve a reliable $YHJ$ spectrum.       
      
    \begin{figure}
     \centering
        \includegraphics[width=.95\columnwidth]{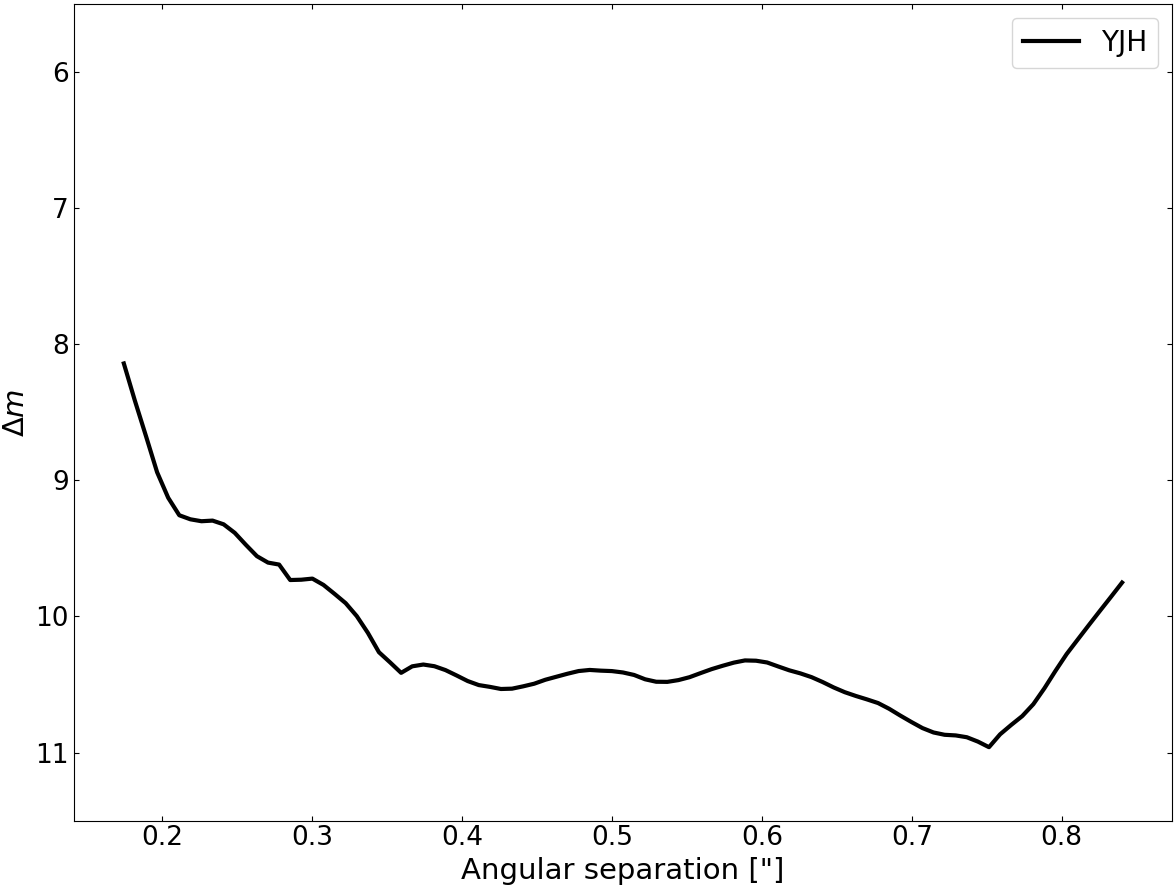}
        \includegraphics[width=.96\columnwidth]{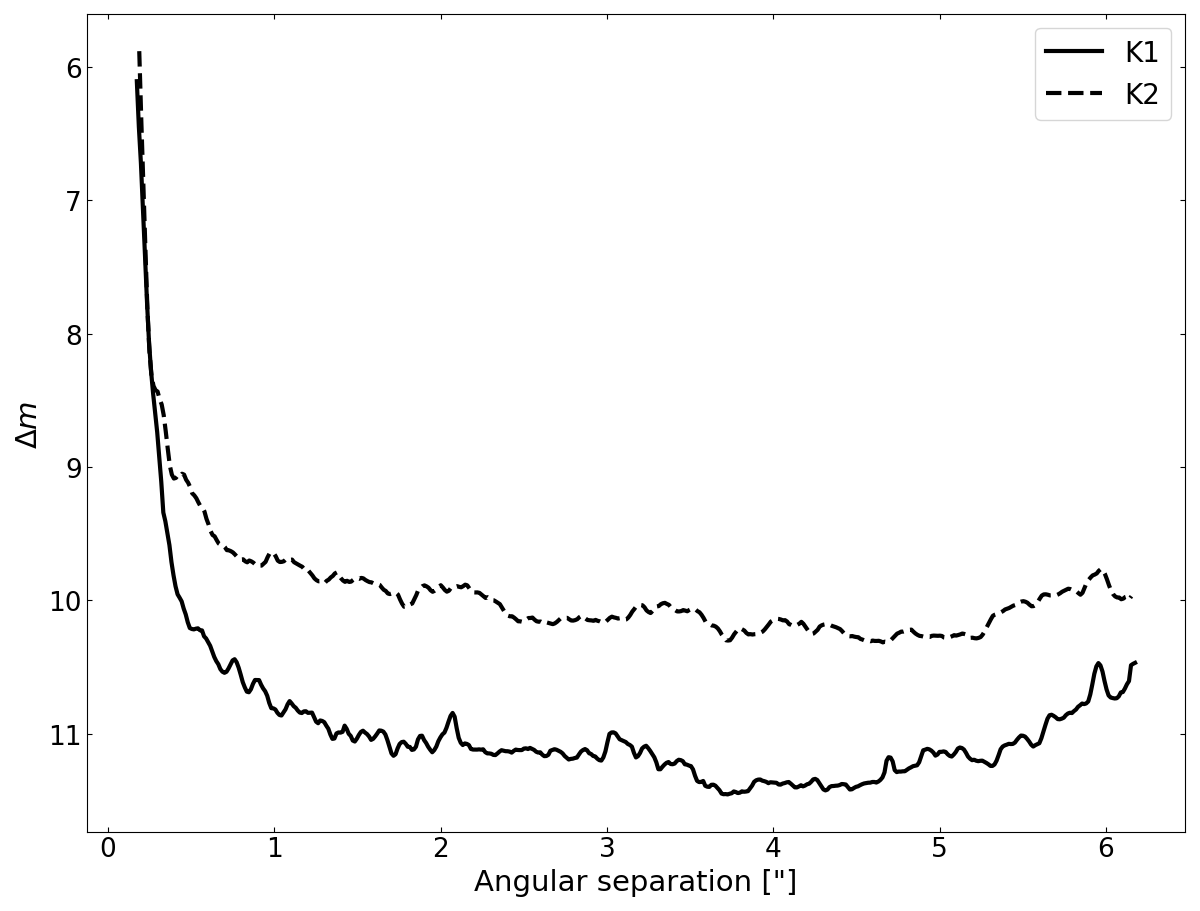}
 \caption{IFS (top) and IRDIS (bottom) detection limits. The contrasts are given in the $YJH$-band for IFS and $K$-bands for IRDIS. The increase in IFS contrast at angular separations $> ~0\farcs75$ result in the presence of artefacts in most images at those separations.}
 \label{f:contrast}
    \end{figure}
      
    \subsubsection{IRDIS sources} \label{s:irdis}
    
    Sources will now be discussed separately.

     \paragraph{\it Tr14-1 ($\equiv$\ HD~93129)} is a visual system with two known O-type components, Tr14-1A ($\equiv$\ HD~93129A) and Tr14-1B ($\equiv$\ HD~93129B) separated by 2\farcs8 \citep{1998AJ....115..821M,2014ApJS..215...15S}. Tr14-1A itself is a O2~If+O3III interferometric binary \citep{2004AJ....128..323N} with an orbital period of the order of a century and a high eccentricity \citep{2014ApJS..215...15S, 2017MNRAS.464.3561M} while Tr14-1B is a O3.5V presumably single star \citep{2014ApJS..215...15S,2014ApJS..211...10S}.
        With an O2~If primary, Tr14-1A is the earliest and hottest O-type star known so far in the Milky Way \citep{2003gafe.conf...17W,2014ApJS..211...10S,2019A&A...621A..63G}. Our data does not allow us to resolve Tr14-1A (it had a separation of a few mas at the time of the SPHERE observations). 
        
        We obtained two separate SPHERE observations centred on Tr14-1A and on Tr14-1B respectively, resulting in partly overlapping FoVs. The observations of Tr14-1A were subject to poor atmospheric conditions and variable absorption due to changing cloud coverage during the observing sequence. This resulted in variable measured fluxes of the detected sources for each rotation frame (which form a time series as described in Sect.~\ref{s:data}). By measuring the flux change for sources at different radial distances, we also noticed that the flux ratio from the first rotation to the last was different for stars close and distant to the central star. That ratio was smaller for closer stars and larger for stars at the edge of the frame. That radial dependence was also seen for sources $180^{\circ}$ from each other (at the opposite side of the image). This change in flux with time was further seen in the PSF frames. Overall, these conditions hindered our ability to directly measure accurate fluxes from the Tr14-1A images. 
        To mitigate these effects, we computed the flux ratios of sources in the Tr14-1A FoV with respect to companion E, observed simultaneously through the Tr14-1A acquisition sequence, and in good weather conditions in Tr14-1B frames. This procedure essentially replaced the PSF contrast by a contrast flux with respect to E. While this allows us to correct for temporal fluctuations in the atmospheric absorption, it cannot correct for its variable spatial component as described above. The limiting detection curves reached a magnitude contrast of $\Delta m > 11.5$ at 2" for the A observations and a magnitude contrast of $\Delta m > 13.5$ at 2" for B.
        
         In total, 74 unique sources were identified in the combined Tr14-1A+B FoVs: 41 sources were detected in the Tr14-1A FoV and 61 sources in the Tr14-1B FoV, 28 of which are detected in both FoVs, including companion E. While the positions are likely correct, we warn against precision usage of the reported flux for the Tr14-1A observations. Indeed a  direct comparison of the measured contrast of  the overlapping  sources shows that the measured fluxes in Tr14-1A are systematically fainter than in Tr14-1B. Given this, whenever possible, we only provide flux measurements obtained from Tr14-1B observations.
                
        In the Tr~14-1A FoV, we detected six previously known bright companions \citep[Tr~14-1B to G,][]{ 2014ApJS..215...15S} as well as two possible binaries in sources S20 and S22, whereas we could only detect three in the Tr~14-1B FoV (A, C \& E) as well as two possible low-mass binaries: S30-31, S34-35 \& S49-50.

Although our positions matched those obtained by SMaSH+, our K-band contrast magnitudes differed for the sources detected in the A observations, which is not surprising given the variable observing conditions described above. The agreement is significantly better with the B image (e.g., CHIPS: $\Delta K_{1,2}(E) = 6.32 \pm 0.01$, SMaSH+: $\Delta Ks(E) = 6.55 \pm 0.05$), also taking into account the fact that the observing bands are slightly different in both  surveys and that the flux calibration strategy differs too.

        \begin{table}[t]
        \caption{Best-fit parameters (Cols.~2-4) of an EEF87's profile (Eq.~\ref{eq:eef87}) adjusted to the MAD data \citep{2010A&A...515A..26S} as a function of the maximum $K_\mathrm{S}$ magnitude considered in the sample (Col.~1). Only values for a $K_\mathrm{S}$ steps of 1.0 are provided for illustration purposes but we have performed and used steps of 0.1 in our analysis.}
            \centering
            \begin{tabular}{c c c c}
            \hline \hline
            $K_\mathrm{S}$ range & $\Sigma(0)$ & $a$ & $\gamma$ \\
                                 & (arcmin$^{-2}$) & arcmin\\
             \hline
             12.0 & 427 &0.04 & 1.5 \\
13.0 & 161 &0.35 & 3.0 \\
14.0 & 281 &0.53 & 3.7 \\
15.0 & 551 &0.73 & 4.8 \\
16.0 & 823 &0.78 & 4.8 \\
17.0 & 947 &1.04 & 6.3 \\
18.0 &1065 &1.50 &10.4 \\
19.0 &1206 &1.48 &10.4 \\
20.0 &1220 &1.48 &10.4 \\
             \hline
             \end{tabular}
            \label{table:EEF87}
        \end{table}

  \paragraph{\it Tr14-2 ($\equiv$\ HD~93128)} is an O3.5V spectroscopic binary \citep{2014ApJS..211...10S}. One previously known visual companion was detected in our SPHERE observations (HD~93128C), alongside 30 new S sources. HD~93128C was measured with $\Delta K = 5.3$mag at $\rho = 3\farcs7$, which again is in good agreement with the measurements obtained in SMaSH+ ($\Delta Ks_\mathrm{C} = 5.4$). HD~93128B, which was detected in SMaSH+ with $\Delta Ks_\mathrm{B} = 2.11$mag at a separation $\rho = 6\farcs55$ lays beyond the SPHERE FoV. Detection limits reached a magnitude contrast of $\Delta m \approx 12$mag at 2".
        
         \paragraph{\it Tr14-8 ($\equiv$\ CD$-$58$^{\circ}$3529, LS~1823)} is an O7V star \citep{2014ApJS..211...10S}. The close-by companion Tr14-8B detected in IFS (see Sect.~ \ref{s:ifs}) has an average K-band contrast of $\Delta K = 3.92$ mag at $\rho = 0\farcs26$. We further detected 35 other S sources. Tr14-8B is so close to the central object that it is present in the PSF images and, to avoid any interference in the PSF modelling, we masked it as discussed in Sect.~\ref{s:reduction}. Detection curves reached a magnitude contrast of $\Delta m \approx 11.5$ mag at 2".
        
          \paragraph{\it Tr14-9} is an O8.5~V star \citep[][]{2014ApJS..211...10S}. We obtained two epochs of observation in ESO periods 98 and 102, i.e., separated by about 2 years. Poor weather conditions in P98 hindered our ability to retrieve accurate fluxes in the same way as for Tr14-1A. Fortunately, observations made in P102 were executed in good conditions and we focus on the latter for contrast measurements. 47 sources were detected in the IRDIS FoV with a particularly bright one close to the source described in Sect.~\ref{s:ifs}. We compared the positions of the sources across the two epochs to check for any movement but no significant displacement could be detected across the two years of observations. Detection curves reached a magnitude contrast of $\Delta m > 10.5$ mag at 2". 
          
         \paragraph{\it Tr14-20 ($\equiv$\ CD$-$58$^{\circ}$3526, LS~1814)} is an O6~V star in the outskirts of the cluster \citep{2014ApJS..211...10S}. This is the least densely populated field, with only 7 S sources detected for this star. The PSF had an unusual shape which caused some issues with spectrum extraction. This PSF was replaced with a scaled version of that  of Tr14-1A to preserve the flux. Limiting magnitude curves reached a contrast of $\Delta m > 10.5$ mag at 2". 
        
        \paragraph{\it Tr14-21 ($\equiv$\ ALS~15207)} is an O9~V spectroscopic binary \citep{2014ApJS..211...10S}. 18 S sources were detected. Detection limit curves reached a magnitude contrast of $\Delta m > 10.5$ mag at 2".

    \begin{table}
    \caption{Summary of companion detections for each FoV. Column 2 gives the total number of detected sources. Columns~3 and 4 give the number of these that satisfies  $P_\mathrm{spur} < 0.2$ and the number of known but unresolved companions, that is, those with a angular projected separation smaller than the resolution capability of  SPHERE ($\rho < 0\farcs15$). 
    }
    \centering
    \begin{tabular}{l c c c}
    \hline\hline
    \vspace*{-3mm}\\
   Central & N. Detected & N. Companion & N. Companion \\
   Object    & Companions  &  $P_\mathrm{spur} < 0.2$ &  $\rho < 0\farcs15$ \\
    \hline
       Tr14-1A  & 41 & 1 & 1 \\
       Tr14-1B  & 61 & 0 & 0 \\
       Tr14-2   & 30 & 1 & 1 \\
       Tr14-8   & 36 & 4 & 0 \\
       Tr14-9   & 47 & 2 & 1 \\
       Tr14-20  &  7 & 0 & 0 \\
       Tr14-21  & 16 & 5 & 1 \\
    \hline
    \label{table:SumComp}
    \end{tabular}
    \end{table}
    
\section{Results and discussion} \label{s:results}

  \subsection{Probabilities of spurious association} \label{s:Pspur}
  
    Detecting sources in an image does not prove that they are gravitationally bound to the central object. Measurements of relative proper motion could inform us on the status of their connection to the central star as it was the case for our observations of Tr14-9, but in a cluster such as Trumpler~14, even common proper motion can only test for cluster membership. 
    In such situations, one has to rely on a statistical evaluation of the companionship likelihood  by estimating the probability that the presence of a companion at a given distance from the central source can be explained by the local source density. 
    
        \begin{figure}
        \centering
        \includegraphics[width=.95\columnwidth]{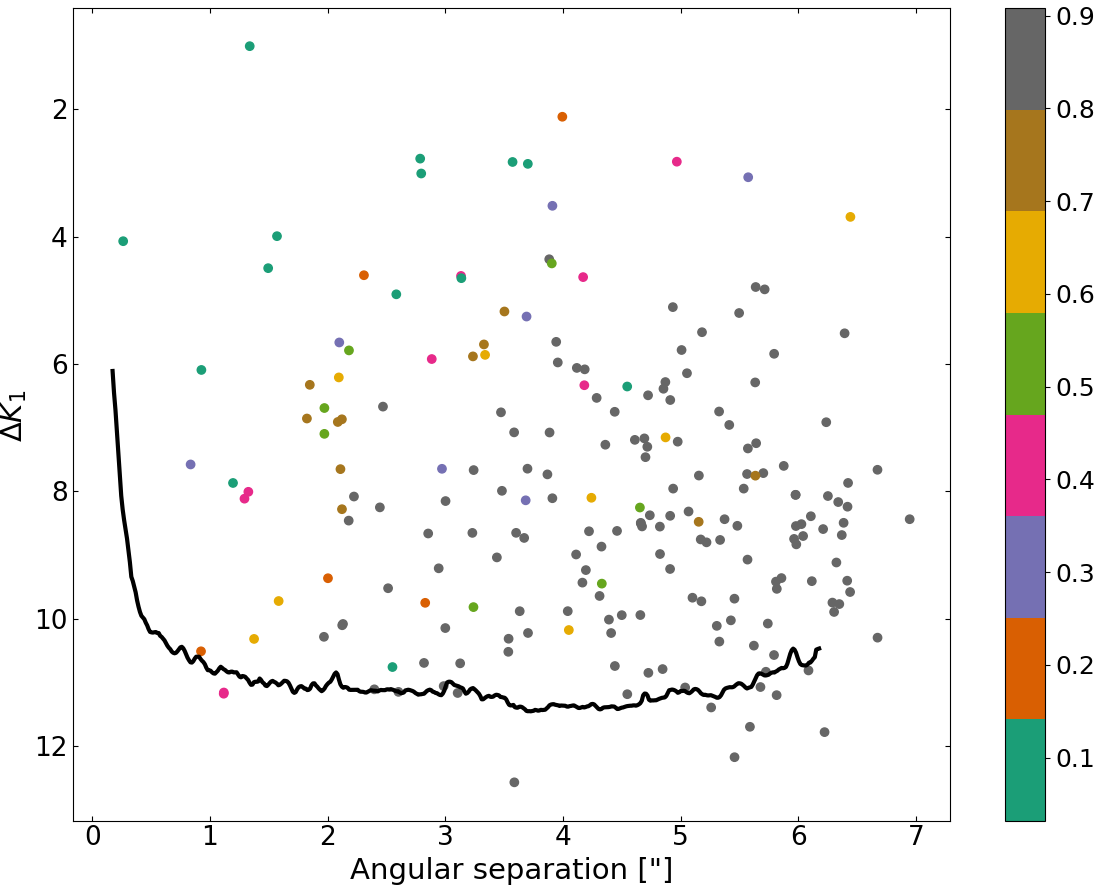}\hspace*{5mm}

        \caption{$K1$-magnitude contrast as a function of radial distance from the central object. Colours indicate spurious alignment probabilities. The figure for $K2$ looks similar and is not included.} 
        
        \label{f:dmag_dist}
        \end{figure}      
        
   In \citetalias{2020A&A...640A..15R}, we estimated the local source density of stars using the VISTA Carina Nebula catalogue \citep{2014A&A...572A.116P}, which  provides a complete infrared survey of stars within $6.7\,\mathrm{deg}^2$ around Eta Car down to a magnitude of $Ks = 19.3$. However, the local density of stars in Trumpler~14 varies significantly at the sub-arcmin scale (see e.g., Fig.~\ref{f:tr14}). Previous studies of  Trumpler~14  \citep[][]{2007A&A...476..199A,2010A&A...515A..26S} have shown that the surface number density of stars as a function of the distance $r$ from the centre of the cluster is well represented by an \citet{1987ApJ...323...54E} profile (EEF87):
    \begin{equation}
        \Sigma(r)=\Sigma_0 (1 + r^2/a^2)^{-\gamma/2}.
        \label{eq:eef87}
    \end{equation}   
Unfortunately, the authors of the previous studies only used two magnitude cuts to compute the density profile, $K_\mathrm{S}<12$ and $<18$. This is insufficient for us and we therefore embarked on re-fitting Eq.~\ref{eq:eef87} applying $K_\mathrm{S}$ magnitude cuts from 12 to 21~mag, by steps of 0.1. In doing so, we noticed that we cannot reproduce the results of \citet{2010A&A...515A..26S}. The surface number density in the cluster core obtained here is about 10 times smaller than that of \citet{2010A&A...515A..26S} for an identical magnitude bin considered. New values of the EEF87 profile parameters are given in Table~\ref{table:EEF87} for representative magnitudes. Armed with these values,
the expected local sources density $\Sigma(r)$ at the location of each of our targets can directly be computed from Eq.~\ref{eq:eef87}. With that information at hand and using a Monte Carlo approach, we then randomly generate 10\,000 populations of stars uniformly distributed over each of our FoVs to determine the probability $P_\mathrm{spur}$ of a source $i$ that at least one star is found at $\rho \le \rho_i$.
  
    For the newly detected companion Tr14-8B, we find that the probability of spurious association is $P_\mathrm{spur} < 1 \times 10^{-3}$, therefore strongly implying its physical association to the Tr14-8 central object. However, such a clearly cut association  is more difficult to state for most other sources as only 13 of them have a spurious association probability  $P_\mathrm{spur} < 0.2$. This results from the very high local source density in Trumpler 14, which is actually high enough to explain most of our detections. Table~\ref{table:SumComp} provides an overview of obtained results for each target, with the number of detected companions in total, those with $P_\mathrm{spur} < 0.2$ as well as known but unresolved companions at $\rho < 0\farcs15$. In total, 13 companions (resolved and unresolved) have a good chance to be physically associated with their central star. Of the 7 stars in our sample, only Tr14-1B (HD~93129B) and Tr14-20 do not have confirmed physical companions so far. Assuming a physical connection likelihood given by $1-P_\mathrm{spur}$ for each detected sources,  one actually obtains an average companion fraction of $2.0\pm0.5$ per star in our sample, where we have assumed Tr14-1A and 1B to be individual objects\footnote{The companion fraction rises to $2.6\pm0.6$ should one assumes that Tr14-1A and 1B are part of the same system.}. This number is identical to that derived from other works \cite[e.g.,][]{2010A&A...515A..26S,2017ApJS..230...15M} but if of course affected by larger uncertainties due to the strong contamination by cluster members in Tr~14.
            
     \begin{figure}
     \centering
        \includegraphics[width=\hsize]{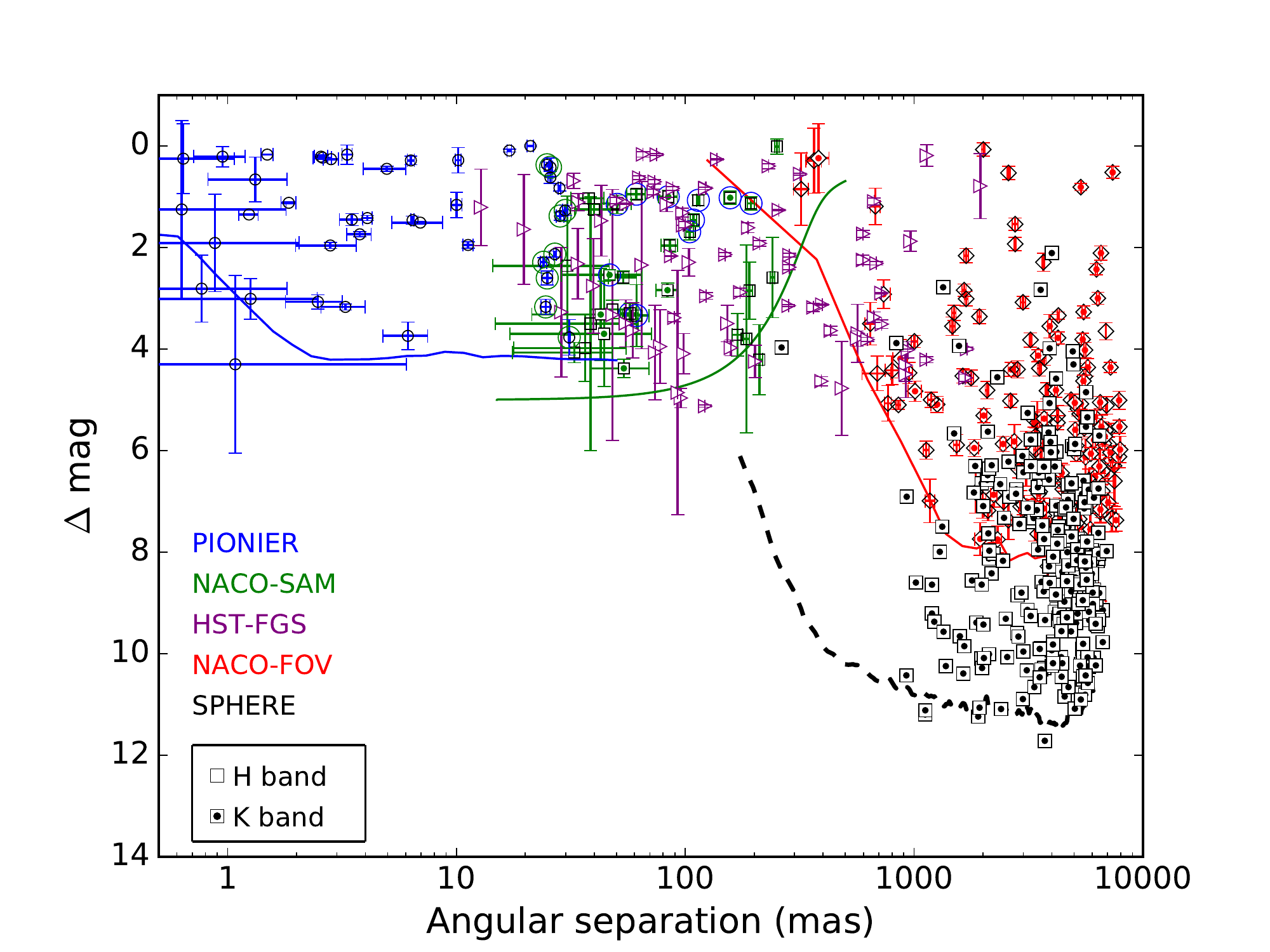}
        \caption{Companion detections in the magnitude contrast vs. angular separation plane showing the Trumpler~14 sources with SPHERE (square symbols), the SMaSH+ \citep{2014ApJS..215...15S} and HST-FGS \citep{2015AJ....149...26A} sources detected. The thick lines give the limiting Detection curves of the different instruments (see legend).}
        \label{f:SMaSHTr14}
        \end{figure} 
        
    \subsection{Spatial distribution of detected sources} \label{s:Dist_source}
    
        We detected 211 unique sources across seven FoVs using the two K-band channels $K1$ ($2.110\,\mu$m) and $K2$ ($2.251\,\mu$m) provided by the IRDIS instrument. Figure~\ref{f:dmag_dist} presents the $K1$ magnitude contrast distribution of the detected sources as a function of the angular separation from their respective central stars. Probabilities of spurious association $P_\mathrm{spur}$ are shown as a colour gradient. As expected, source detected at close ranges and brighter sources have lower probabilities of being spurious companions than fainter sources and sources at the larger end of the angular separation range.
        
\begin{figure}
         \centering
        \includegraphics[width=\hsize]{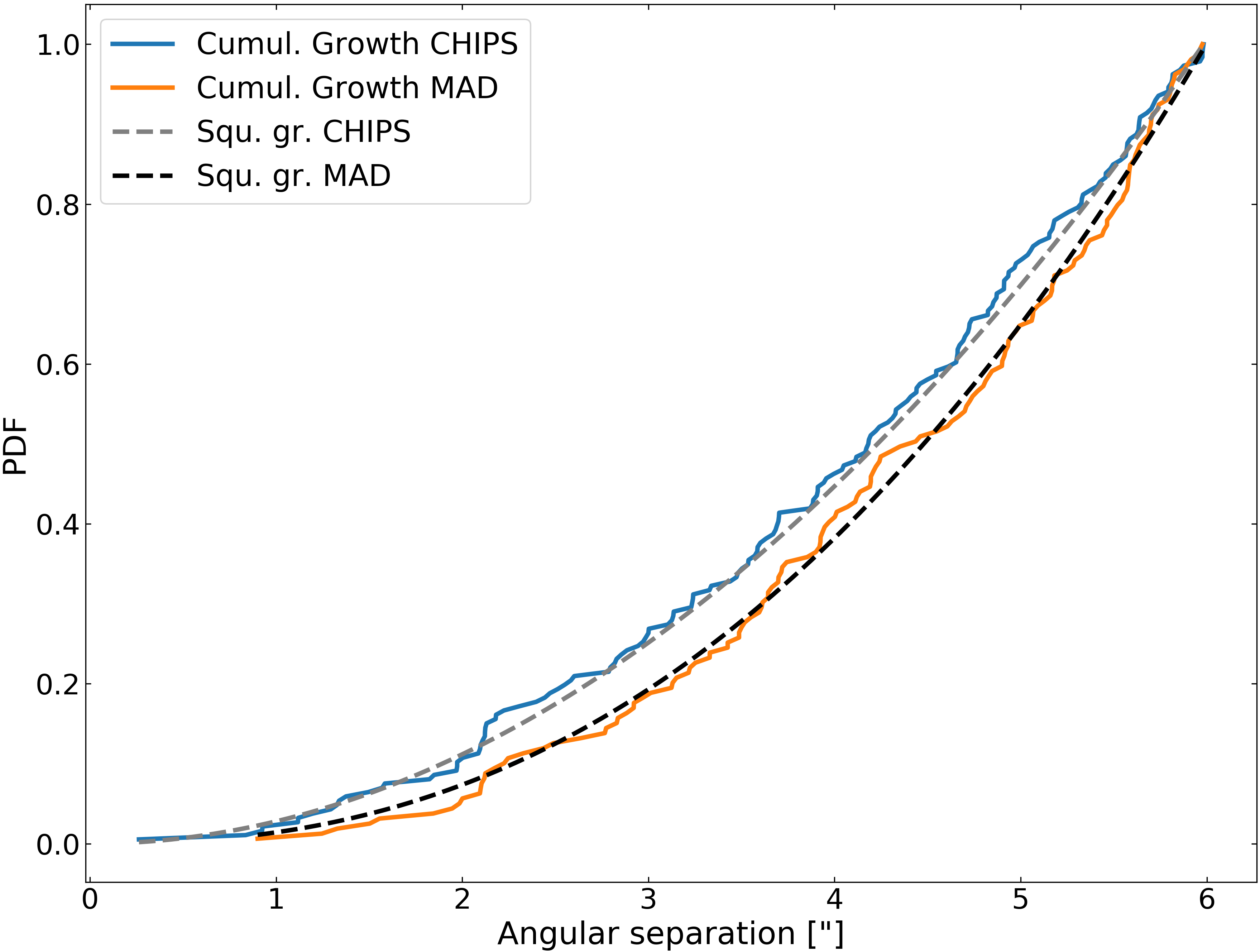}
        \caption{Growth  curves of the detected sources in CHIPS (blue line) and MAD (orange line) for the Trumpler~14 cluster with respect to the angular separation of the central star, limited to 6". A squared growth (grey dashed line for CHIPS, black for MAD) is overlaid.}
        \label{f:growthc}
        \end{figure}

        We can directly compare our results with previous high-angular resolution surveys such as the SMaSH+ and the HST-FGS surveys \citep[Fig.~\ref{f:SMaSHTr14},][]{2014ApJS..215...15S, 2015AJ....149...26A}. From these results and the ones first obtained with QZ Car \citepalias{2020A&A...640A..15R}, it is clear that SPHERE opens a window in the parameter space that was previously unreachable. It allows for the first time to look at closer separations for the fainter visual companions which  correspond to the low-mass end of the companion mass function of massive stars. We also observe an aggregation of sources, a phenomenon previously seen in SMaSH+ and the QZ Car system, at angular separations larger than 2" (i.e. projected physical separation $>4.6 \times 10^3$ au). This raises the question of whether this larger number of distant companions is itself a consequence of the larger effective surfaces of the images at large distance or whether there is a genuine lack of companions at short separation. In this respect, the growth curve  in Fig.~\ref{f:growthc}, shows that the cumulated number of companions as a function of the separation follows a $\rho^2$  distribution, hence favouring in the present case the first option. This is expected for Trumpler~14 as we have shown earlier that the number of sources is driven by the local source density of the cluster's  profile.

    \subsection{Comparison with previous AO imaging} \label{s:MAD}

\begin{figure}
         \centering
        \includegraphics[width=\hsize]{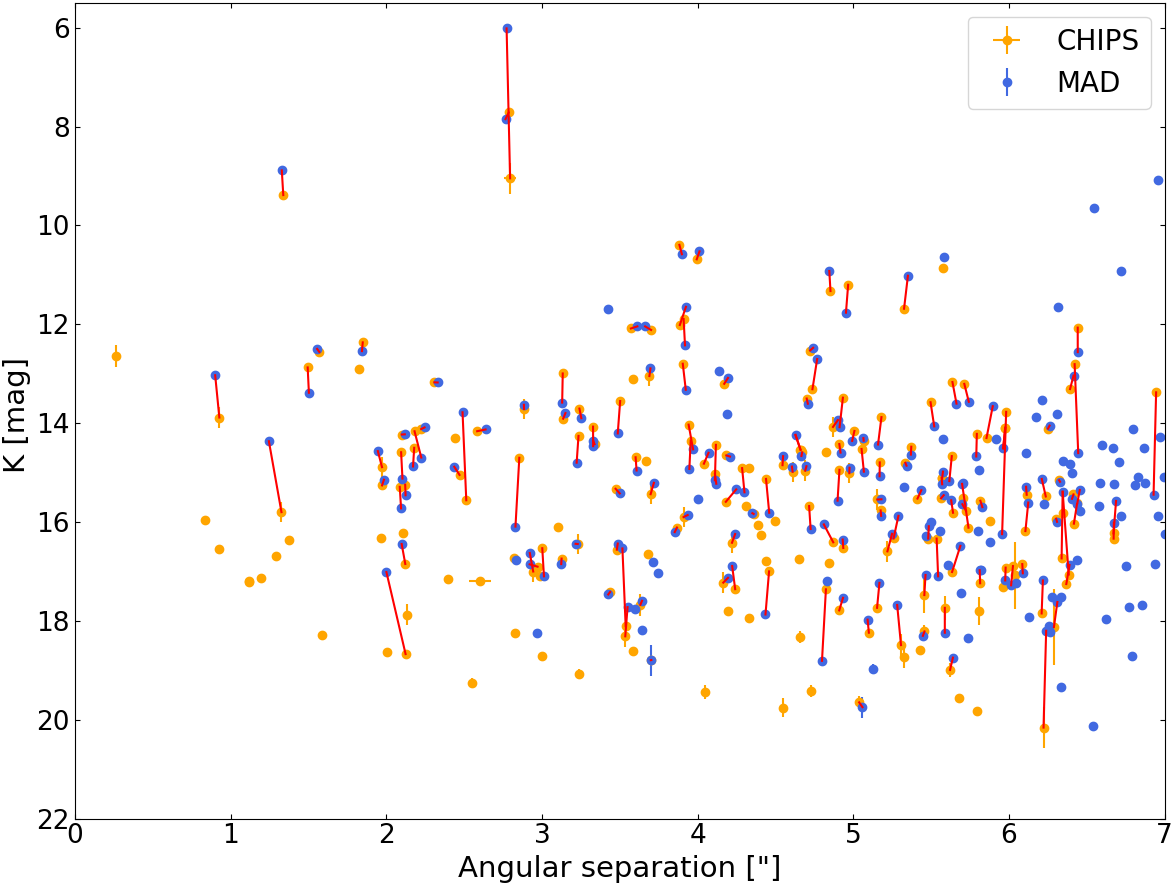}
        \caption{K-band magnitude distribution of detected sources in the Trumpler 14 cluster with the CHIPS (orange; $K_1$) and MAD (blue;$K_\mathrm{S}$) samples as a function of angular separation from the central star. Sources within 0\farcs1 from one another are connected by a solid red line. MAD sources brighter than $K = 8$ are likely impacted by non-linearity and/or saturation effects impacting the MAD observations. }        
        \label{f:Kmadchips}
        \end{figure}

   The  Trumpler~14 cluster has been one of a limited few scientific targets of the Multi-conjugate Adaptive optics Demonstrator \citep[MAD,][]{2003SPIE.4839..317M}. Using three natural guide star (NGS) across the FoV, MAD was able to provide improved AO-corrected images over a wider field than single NGS-AO observations. The Trumpler~14 MAD campaign covered a 2\arcmin$\times$2\arcmin\ FoV in five dithering positions in both $H$ and $K_\mathrm{S}$ band \citep{2010A&A...515A..26S, 2011MNRAS.418..949R}. 
    We retrieved positions, $H$ and $K_\mathrm{S}$ bands  magnitudes for every source detected from the catalogue of \citet{2010A&A...515A..26S} and compared the performance of MAD and SPHERE close to the massive stars.  Within 6\arcsec\ of each O star in our sample, CHIPS detected 185 sources whereas MAD detected 159. Figure~\ref{f:Kmadchips} shows that SPHERE is able to detect sources at fainter magnitudes and at closer separations than MAD could (about 3~mag fainter at 1\arcsec\ and about 1~mag  at 2~\arcsec). As discussed in \citet{2010A&A...515A..26S}, the MAD close-companion detection capability drops below 60\%\ for  bright central objects ($\rho < 2$\arcsec, $K<9$) and indeed SPHERE detects 36 companion in that range while MAD only detect 20 below 2\farcs5. Figure~\ref{f:Kmadchips} compares the properties of the companion detected near our sample stars in both surveys, where one can see again excellent agreement of the separation and  deviations of the order of 0.3~mag (rms) in the retrieved magnitudes which, as discussed previously, likely originate from the different bands between the two instruments and from their different calibration strategy. 
        
        \subsection{Radial distribution of companions}
        
\begin{figure*}
        \centering
        \includegraphics[width=.97\columnwidth]{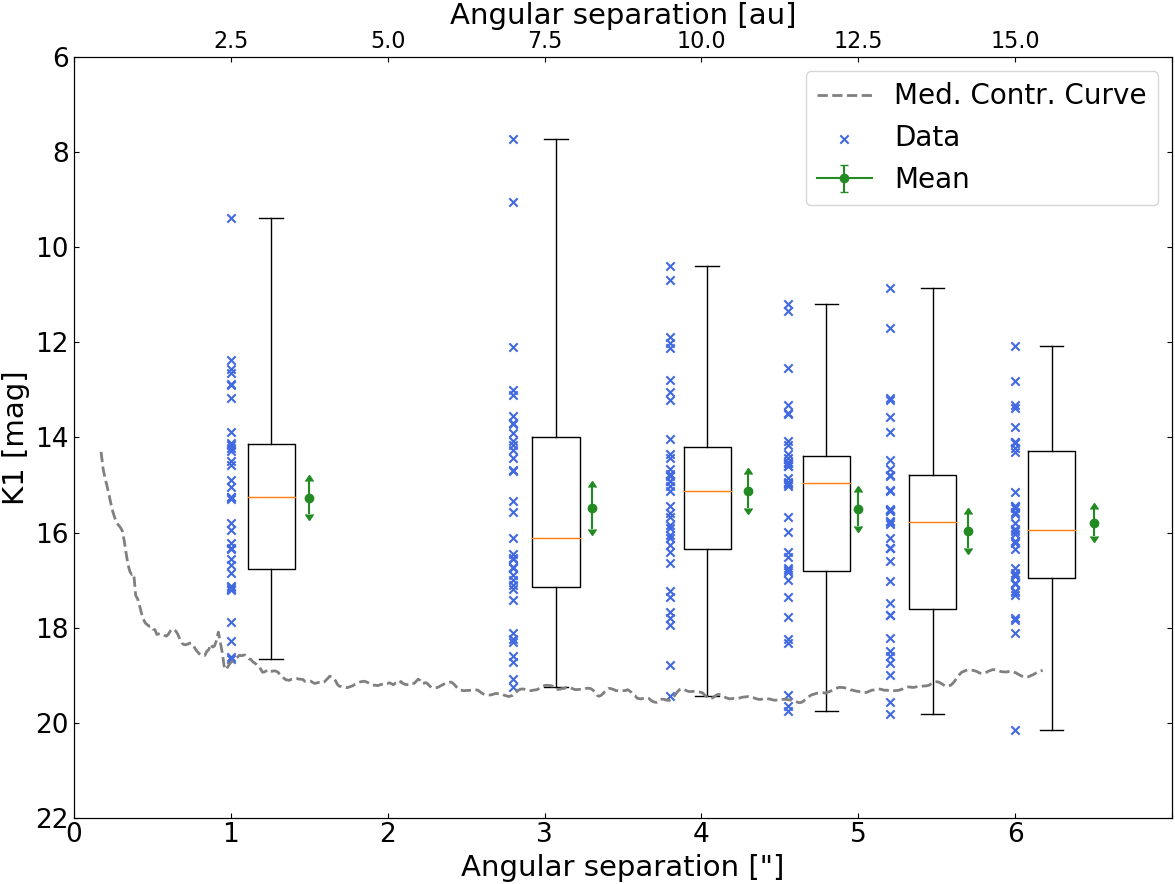}\hspace*{5mm}
        \includegraphics[width=.97\columnwidth]{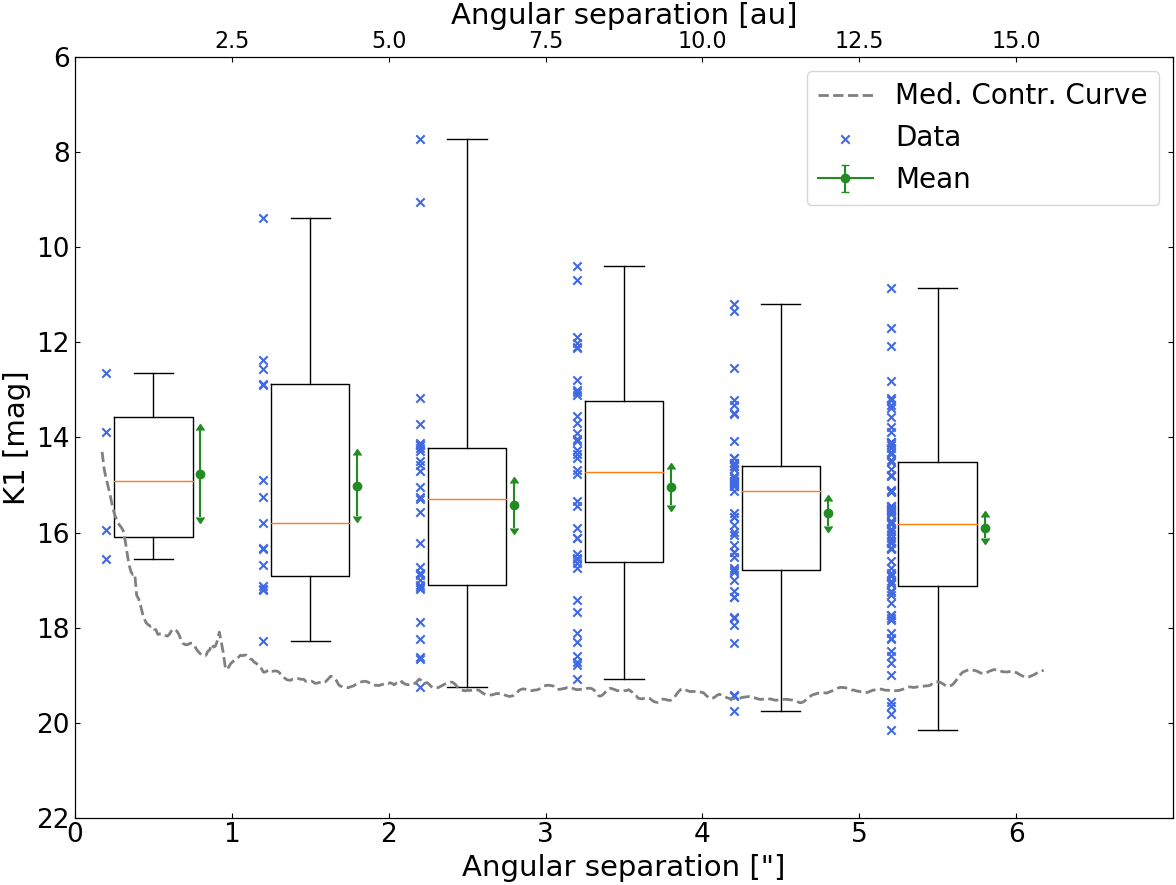}
        \caption{Box plots of the distribution of K1 magnitudes of detected sources as a function of radial distance from the central stars. Boxes mark the range between the 0.25 and 0.75 percentiles, with orange lines indicating the median magnitude for each bin. The mean and its 1$\sigma$-uncertainty is provided to the right-hand side of each box (green dot and arrows) while the $K$-mag distribution of the stars in each bin is indicated to the left-hand side of each box (blue crosses). The extent of the sample is provided by the vertical bars. The dashed line provide the IRDIS median contrast curve. The 211 sources were divided into six bins, but the choice of the bins differ in each panel. \textit{Left panel}: Each bin has an equal amount of sources, 35 sources per bin. \textit{Right panel}: The IRDIS FoV was divided into equal-sized bins. The former plot alleviates the expected $r^2$ dependence of the effective surface of equally-spaced bins (hence of the expected number of sources) as well as bin-to-bin variations due to difference sample sizes. The latter plot allows to go closer to the central star.}
        \label{f:boxplots}
        \end{figure*}

        As previously done in Fig.~\ref{f:dmag_dist}, we can look at the distribution of absolute magnitudes for all sources as a function of radial distance from the central star, represented as box plots in Fig.~\ref{f:boxplots}. No obvious trend is visible, with most of the 0.25 to 0.75 percentile intervals (the boxes)  overlapping significantly. Two features are worth noting, although their statistical significance is difficult to assess: the average magnitude in the bin furthest away from the central stars is slightly but systematically lower than the innermost bins. Second, the brightest companions with $K<10$ mag (hence likely late-O /early-B stars) are all within the inner 3\arcsec\ (approx. $7.5\times 10^3$ au) from their  central object. There is a lack of faint/low-mass stars in the inner arcsec, but this can result from the smaller surface cross-section and lower limiting contrast in the IFS FoV. Similarly the faintest stars are found beyond 4\arcsec, which is hard to explain by observational limitations as the limiting contrast curves tend to degrade at the largest separation (see e.g., Fig.~\ref{f:contrast}). If truly belonging to Trumpler 14 however, those faintest objects are beyond the stellar mass regime and likely well into the brown dwarf domain.
                
    \subsection{Physical properties of the detected sources} \label{s:phys_source}
    
        The physical properties of the one companion detected in the IFS FoV have been obtained by fitting the $YJH$-SED in Sect.~\ref{s:ifs}. Here, we further attempt to constrain the physical properties of the IRDIS sources, namely their masses and ages, following the same procedure as in \citetalias{2020A&A...640A..15R}. We acknowledge that a fully open fit is largely degenerate, this is why we restrict the fit of the physical properties to  combinations of parameters that are supported by evolutionary models (see discussion in Sect.~\ref{s:data}). We further adopted a fixed distance of 2.5~kpc \citep{2004A&A...418..525C,2019ApJ...870...32K} and we assume that there is no differential reddening between the sources detected in a given field and its central object. Using the same flux calibration method as  described in Sect.~\ref{s:ifs} for Tr14-8B, the  $K1$ and $K2$ contrast are converted to absolute fluxes using an appropriate FASTWIND model for the central object (adopting the physical parameters listed in Table~\ref{table:fastwind}  and galactic metallicity). $\chi^2$ maps were then obtained by comparing these fluxes to LTE  ATLAS9 and PHOENIX atmospheric models. ATLAS9 and PHOENIX both cover stellar parameters such as the effective temperature $T_{\mathrm{eff}}$, the surface gravity $\mathrm{log}g$, the radius, the mass, age and luminosity. The addition of the PHOENIX models \citep[][]{2013A&A...553A...6H} compared to Paper~I allows us to probe cooler temperatures  than the lower $T_\mathrm{eff}$ limit  of the ATLAS grid at around 3500~K. PHOENIX's lower temperature limit reaches $T_\mathrm{eff} = 2300$~K, so that the models can reach pre-main sequence stars down to $0.1 M_{\odot}$.
                
\begin{figure*}
        \centering
        \includegraphics[width=\columnwidth]{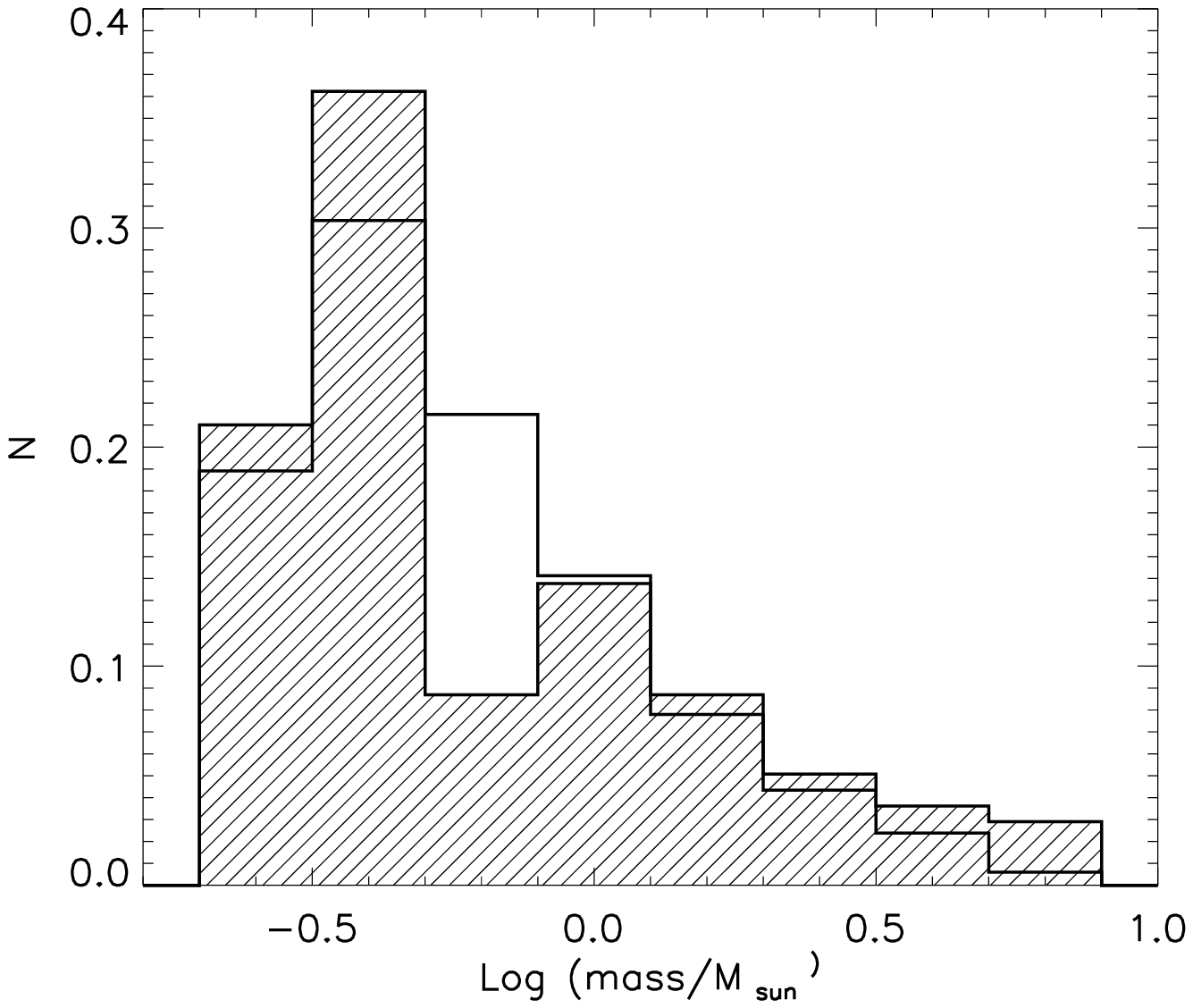}
        \includegraphics[width=\columnwidth]{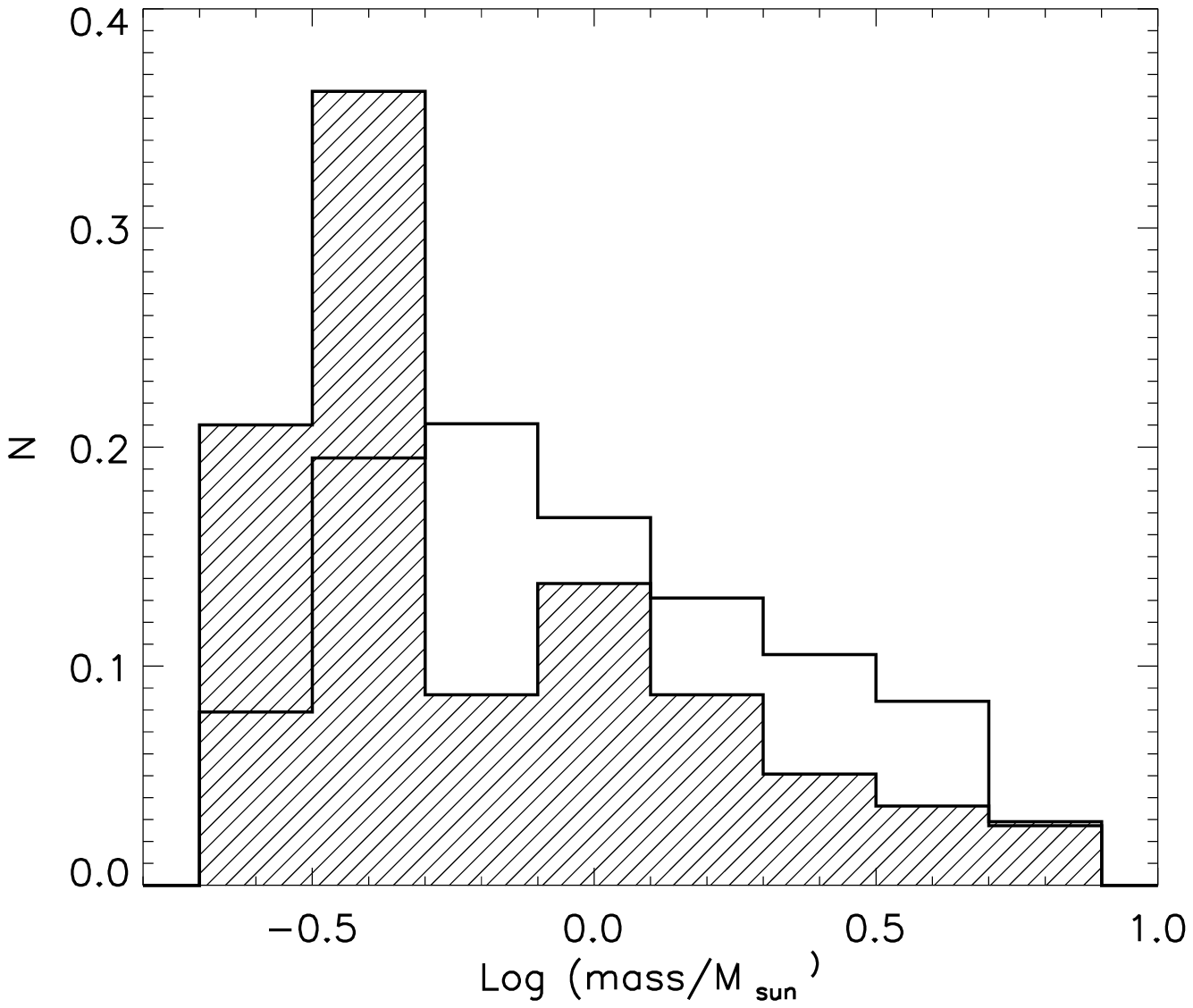}
        \caption{\textit{Left:} Comparison of the normalised distribution of fitted masses above 0.25 $M_{\odot}$ (filled histogram) and a random stellar population drawn from the \cite{2003PASP..115..763C} field IMF (empty histogram). \textit{Right:} Comparison of the normalised distribution of fitted masses above 0.25 $M_{\odot}$ (filled histogram) and a random stellar population drawn from the Tr14 mass distribution from \cite{2011MNRAS.418..949R} (empty histogram).}
        \label{f:IMF comparison}
        \end{figure*}

        Since there is a degeneracy between the age and the mass (see discussion in Paper~I), we only allowed for solutions in the age range 0.2-10 Myrs, accommodating for the expected age from previous findings of Tr14-8B, while also taking into account an older population of stars which could fit into the overall age of the Carina region. The procedure was repeated for both the ATLAS9 and PHOENIX grid, therefore obtaining two different best-fit models for every source. The final adopted model is the one with the lowest $\chi^2$ that also fits within the age limits set. A majority of PHOENIX models provided better fits compared to ATLAS9, indicating that most sources are lower-mass stars. The best-fitting solution implied that 99 sources are younger than 1 Myr and 103 are older.
        32 sources are found with a mass of 0.1~$M_{\odot}$, the low-mass limit of the underlying evolutionary grid. This indicates that they might be of even lower mass or that they are background objects with similar $K$-band magnitude and colours as an 0.1~$M_{\odot}$ PMS stars in Tr~14. The latter explanation is supported by the fact that this cluster is located at Galactic latitude $b = -0.6\degr$, that is very close to the Galactic plane where background stars are most likely subject to higher extinction. In either cases, their best-fit $\chi^2$ remains within the acceptable range within the age limits set. The distribution of retrieved masses above 0.25~$M_{\odot}$ is displayed in Fig.~\ref{f:IMF comparison}.

        \subsection{The mass function in the vicinity of massive stars}
    Using the results of the previous section, we constructed the mass function of the sources detected by SPHERE within the FoV of each central massive star in Tr~14.
        In doing so,  we restricted our analysis of the mass distribution to masses above 0.25$M_{\odot}$, in line with the analysis performed   by \cite{2011MNRAS.418..949R} for the Tr14 region. This allowed us to directly compare our results to previous studies without being significantly affected by the limitations of our model grid.
        We tested two possible shapes of the mass functions. First, we considered a field initial mass function (IMF) which is composed by a log-normal distribution following \citet{2003PASP..115..763C} for masses below 1~$M_{\odot}$ and a \citet{1955ApJ...121..161S} power-law above 1~$M_{\odot}$.
        Secondly, we adopted a broken power-law with break-point at 0.53~$M_{\odot}$ which, according to  \citet{2011MNRAS.418..949R}, is representative of the mass function of low-mass stars in Trumpler~14. We then run a Kolmogorov?Smirnov (KS) test between the observed mass distribution in the vicinity of the massive stars and each of the two mass functions described above.
      We obtained KS probabilities of the order of $5\times 10^{-3}$ for the comparison with \citeauthor{2003PASP..115..763C} and of $2\times 10^{-12}$ with \citeauthor{2011MNRAS.418..949R}, indicating that we can confidently rule out the possibility that they are drawn from the same parent distribution.
        By having a closer look at the distribution of masses in the vicinity of massive stars in Tr~14 compared to those derived by \cite{2003PASP..115..763C} in the galactic field and by \cite{2011MNRAS.418..949R} in the Tr14 region, one notices that, despite the low KS probability value, the peak of our mass function in the  0.3 to  0.5~$M_{\odot}$ mass range matches with the one from Chabrier (field). Compared to the Rocheau (Tr14) mass function, however, we see a lack of sources with masses between 0.5 and 0.8~$M_{\odot}$, indicating a possible influence of massive stars on the local IMF (see Fig.~\ref{f:IMF comparison}).

        As mentioned earlier, our flux calibration relies on theoretical assumptions and may introduce a small shift in the derived absolute fluxes, and of the star masses. Yet, the comparison with the MAD data in Fig.~\ref{f:Kmadchips} shows that, on average, there is little difference between the CHIPS and MAD so these cannot explain the systematic shift in the mass function peak towards higher masses. Alternatively, infrared excesses due to the presence of a disk would impact our mass determination, leading to a slightly overestimated mass. Without access to multi-band photometry for all our sources, it is unfortunately not possible to conclude on the impact of such infrared excess.
        Until now there has been no claims of evidences for an environmental effect due to the presence of massive star on the shape of the low-mass star mass distribution \citep[see e.g.,][]{2021arXiv210108804D}. While our results are preliminary, Trumpler 14 is certainly  one of the best possible laboratories to further investigate this hypothesis.

\section{Summary and conclusions} \label{s:Ccl}

    In this second paper of a series, we presented SPHERE observations of the entire O-star population in the young, compact  cluster Trumpler~14. Using a combination of the IRDIS and IFS instruments, 211 sources were detected in seven $12"\times12"$ FoVs around seven O-type stars. The median detection limit of our observations reaches contrasts better than 9 mag at $0\farcs2$ and better than 11 mag at 2". This enables us to detect sub-solar mass stars in Trumpler~14 over the entire FoV of our observations. 
    
    Over the course of two epochs of observations separated by over 2 years for the detected sources of Tr14-9, we obtain a negligible relative motion, indicating that most detected sources are indeed genuine cluster members.     
    
    The detected sources cover a wide contrast range of $\Delta K \approx 2 - 13$ mag, but the majority  is to be found between $\Delta K = 6$ to 10 mag. The radial distribution of the detected sources is well explained by the local source density in the Trumpler~14 cluster and, indeed most have a large chance of spurious association apart from the brightest and closest sources. No clearly-cut trend could be found in the radial location of the sources, but there is a tentative indication that the most massive companions are found closer to their central star, while the least massive ones are only detected further out.

    We  estimated the physical properties of detected sources by comparing their measured flux to  ATLAS9 and PHOENIX models attached to pre-main sequence evolutionary tracks covering the age range between 0.2 and 10~Myr.  One source, the newly discovered Tr14-8B, is close enough to its central object to fall within the IFS FoV and a full $YJHK$ SED was therefore obtained, indicating that this star is in an 0.5~Myr-old, 1.3~\msun\ pre-main sequence star. The obtained age is in remarkable agreement with previous age estimate for the Trumpler~14 cluster and demonstrate the capability of using pre-main-sequence companion stars to massive stars as an age tag. 
    
    A similar fit to the sources detected in IRDIS yielded mass values between 0.1 (the lower limit of our grid) and 7~\Msun{}. The reconstructed mass function above 0.25~\Msun{} is neither statistically compatible with the field IMF nor with the mass function of the overall Trumpler 14 cluster. The mass function in the vicinity of the O-type stars shows a dearth of masses in the range of 0.3 to 0.6~\msun{}  and a peak at about one solar mass. We discuss possible observational causes that may affect our mass estimate, in particular the potential presence of an infrared excess large enough to bias our mass estimate towards larger masses. Clearly, follow-up observations with SPHERE/IRDIS in the $YJH$-band are needed to clarify this important point.

    Finally, the present results clearly illustrate that SPHERE provides the necessary capabilities to detect and characterise faint companions in the vicinity of massive stars. In a following paper of this series, we will present the remainder of the survey, focusing on the less crowded environment of the overall Carina region. This will alleviate the  possible confusion between chance alignment and genuine companions, hence allowing us to characterise the low-mass end of the companion mass function of massive stars.

\section*{Acknowledgements}

  This work is based on observations collected at the European Southern Observatory under programs ID 095.D-0495(A), 098.C-0742(A) and 0102.C-0104(A). We thank the SPHERE Data Centre, jointly operated by OSUG/IPAG (Grenoble), PYTHEAS/LAM/CeSAM (Marseille), OCA/Lagrange (Nice) and Observatoire de Paris/LESIA (Paris) and supported by a grant from Labex OSUG@2020 (Investissements d'avenir a ANR10 LABX56). We especially thank P. Delorme, E. Lagadec and J. Milli (SPHERE Data Centre) for their help during the data reduction process.

  We acknowledge support from the FWO-Odysseus program under project G0F8H6N. This project has further received funding from the European Research Council under European Union's Horizon 2020 research programme (grant agreement No 772225, MULTIPLES \& No 819155).

  \textit{Facilities:} VLT UT3 (SPHERE)

\bibliography{references}

\begin{appendix} 
  \section{Observing setup and atmospheric conditions} \label{a:Obs}
    Tables~\ref{t:obs_setup} and \ref{t:obs_weather} provide the journal of the instrumental setup, the journal of the observations and the weather conditions during our observing campaign.
    
    \begin{table*}
    \caption{SPHERE instrumental setup for {\sc flux} (F) and {\sc object} (O) data.}
    \label{t:obs_setup}
    \centering
    \begin{tabular}{l c c c c c c c c}
    \hline\hline
    \vspace*{-3mm}\\
        & \multicolumn{4}{c}{IFS} & \multicolumn{4}{c}{IRDIS}  \\ 
    Object & NDIT (O) & DIT (O) [s] &  NDIT (F) & DIT (F) [s] & NDIT (O) & DIT (O) [s] &  NDIT (F) & DIT (F) [s] \\
    \hline
    Tr14-1A & 8 & 8 & 20 & 1.65 & 2 & 8 & 16 & 0.837  \\
    Tr14-1B & 10  & 16 & 8 & 8 & 2 & 16 & 16 &  8     \\
    Tr14-2  & 10   & 16 & 8 & 16 & 4 & 16 & 8 & 16  \\
    Tr14-8  & 10  & 64 &  4 & 16 & 5 & 32 & 8 & 8   \\
    Tr14-9 (P98) & 9  &  32 & 20 & 1.65 & 2 & 32 & 20 & 8 \\
    Tr14-9 (P102) &  10  &  64 & 4 & 16 &  5 & 32 & 8 & 8   \\
    Tr14-20 & 10 & 16 & 16 & 16 & 2 & 16 & 32 & 8 \\
    Tr14-21 &  12  & 64  & 4 & 16 & 3  & 64 & 4 & 16  \\
    \hline
    \end{tabular}
    \end{table*}

    \begin{table*}
    \caption{SPHERE observing conditions}
      \label{t:obs_weather}
    \centering
    \begin{tabular}{l c c c c c}
    \hline\hline
    \vspace*{-3mm}\\
    Object & Obs. Date & Airmass & PAV (\degr) & Seeing & $\tau_{0}$ (ms)\\
    \hline
    Tr14-1A & 11.02.2016 & 1.5 & 3.26 & 1.14 & 3.1 \\
    Tr14-1B & 01.02.2016 & 1.34 & 5.53 &  0.78 & 3.1    \\
    Tr14-2  & 11.04.2015 & 1.225 & 13.82 & 0.81 & 3.3  \\
    Tr14-8  & 19.03.2019  & 1.221 & 4.21 & 0.63 & 7.7   \\
    Tr14-9 (P98) & 04.03.2017 & 1.22 & 9.25 & 1.15 & 5.8\\
    Tr14-9 (P102) &  27.03.2019 & 1.221 & 4.21 & 0.62 & 6.5  \\
    Tr14-20 & 17.03.2016 & 1.226 & 8.86 & 0.88 & 3.4\\
    Tr14-21 &  22.02.2019 & 1.221 & 4.80 & 0.96 & 8.0 \\
    \hline
    \end{tabular}
    \end{table*}

  \section{IRDIS post-processed images} \label{a:SIRDIS_obs}
\begin{figure*}
        \centering
        \includegraphics[width=.95\columnwidth]{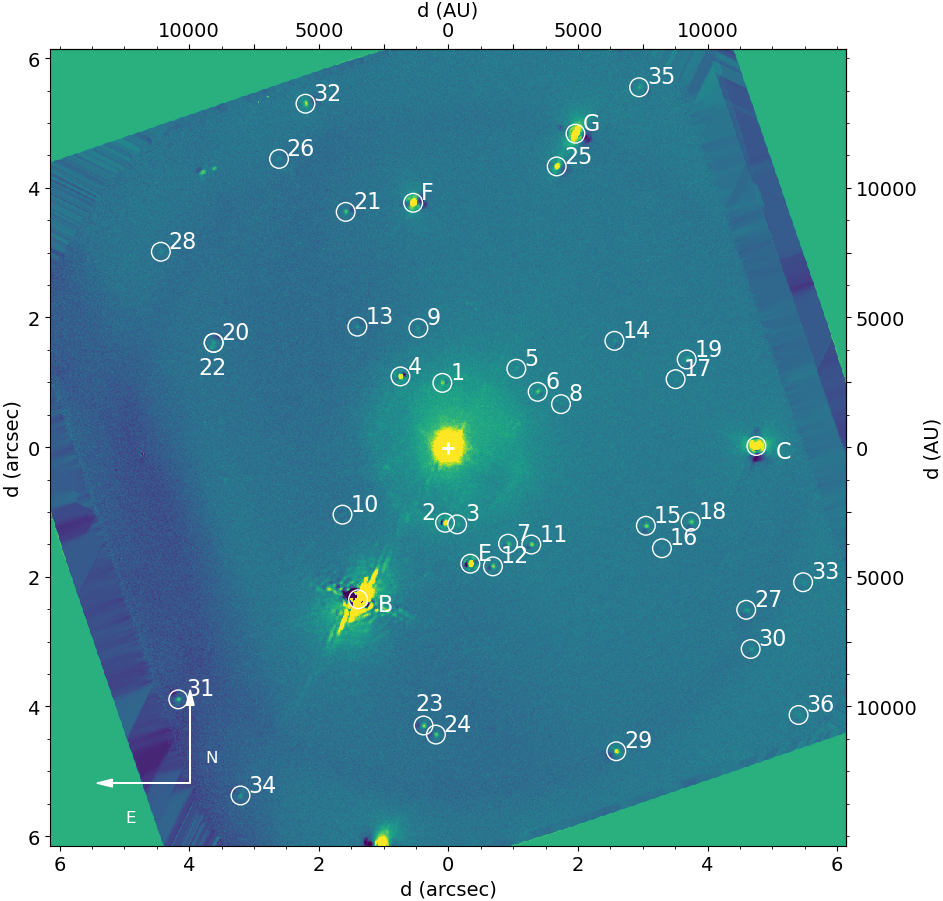}\hspace*{5mm}
        \includegraphics[width=.95\columnwidth]{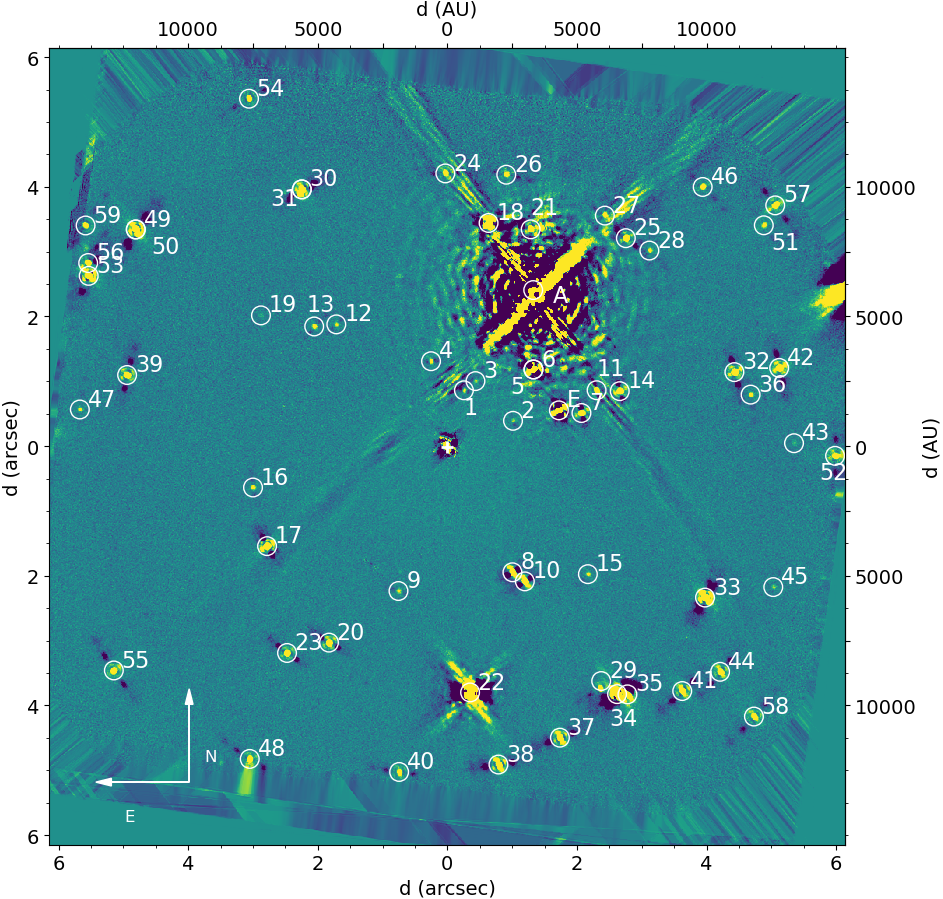}
        \caption{Same as Fig.~\ref{f:CD-583529}. \textit{Left}: Tr14-1A. \textit{Right}: Tr14-1B.}\label{f:irdis_app1}
        \end{figure*}

\begin{figure*}
        \centering
        \includegraphics[width=.95\columnwidth]{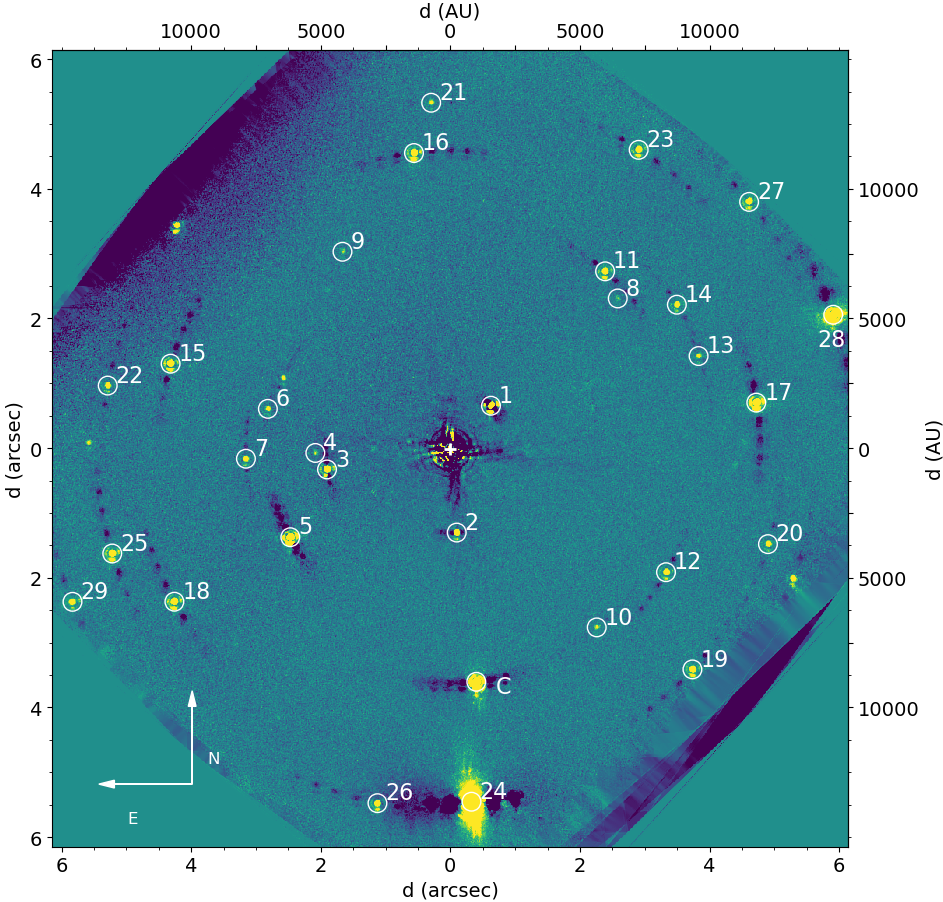}\hspace*{5mm}
        \includegraphics[width=.95\columnwidth]{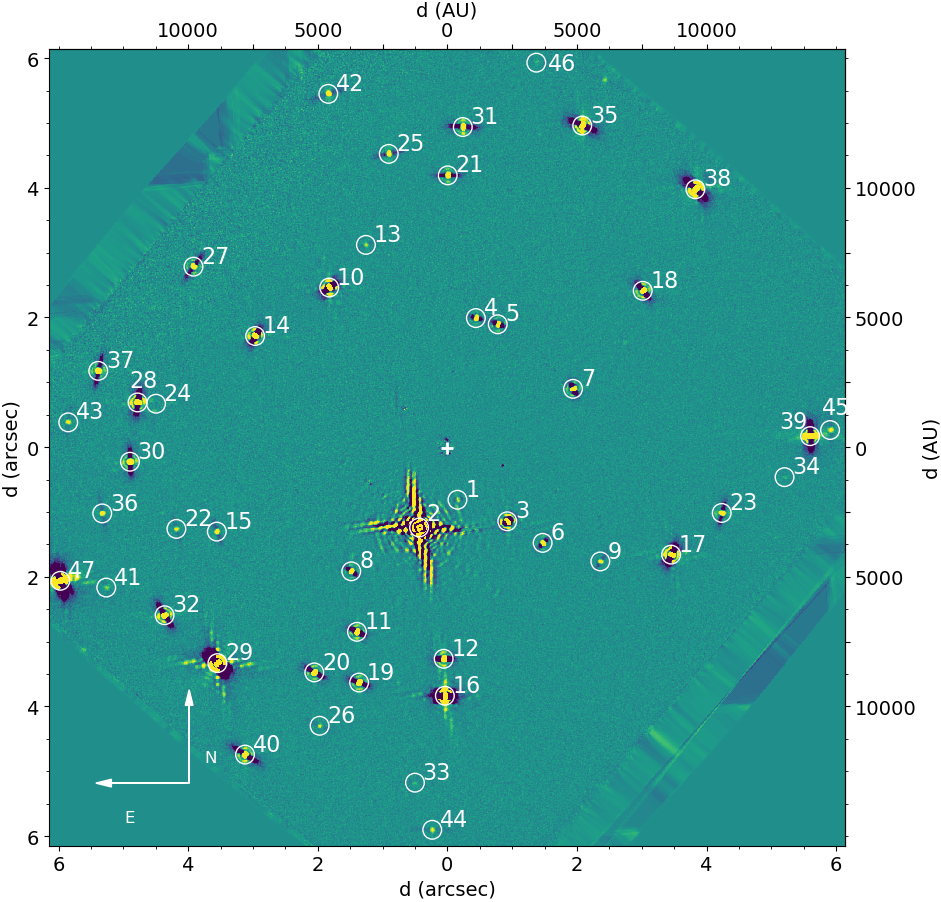}
        \caption{Same as Fig.~\ref{f:CD-583529}. \textit{Left}: Tr14-2. \textit{Right}: Tr14-9.}\label{f:irdis_app2}
        \end{figure*}

\begin{figure*}
        \centering
        \includegraphics[width=.95\columnwidth]{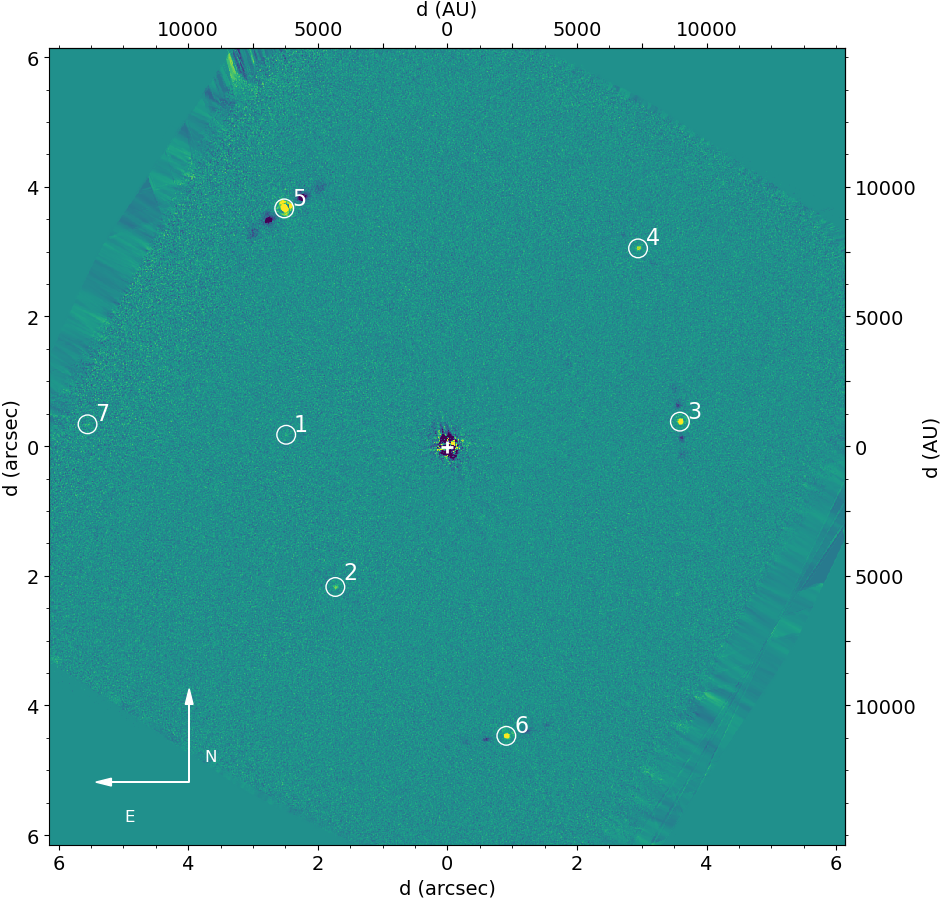}\hspace*{5mm}
        \includegraphics[width=.95\columnwidth]{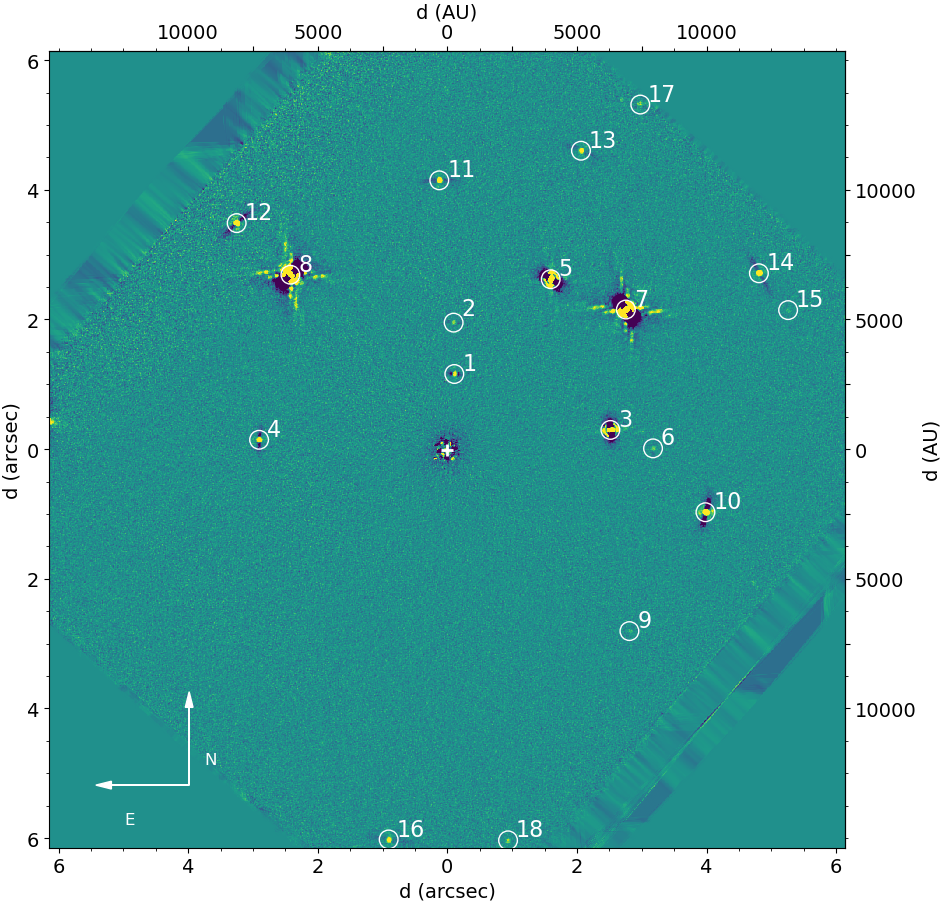}
        \caption{Same as Fig.~\ref{f:CD-583529}. \textit{Left}: Tr14-20. \textit{Right}: Tr14-21.}\label{f:irdis_app3}
        \end{figure*}

        Figures \ref{f:irdis_app1} to  \ref{f:irdis_app3} present the post-processed collapsed images of the IRDIS FoV ($12\farcs \times 12\farcs$) of each of our target with detected companions sources identified and labelled according to the nomenclature of Table~2. The FoV of Tr14-8 is given in Fig.~\ref{f:CD-583529} and is not repeated here.
        
\end{appendix}
\end{document}